\documentclass[pdflatex,sn-basic]{sn-jnl}

\usepackage{graphicx}%
\usepackage{multirow}%
\usepackage{amsmath,amssymb,amsfonts}%
\usepackage{amsthm}%
\usepackage{mathrsfs}%
\usepackage[title]{appendix}%
\usepackage{xcolor}%
\usepackage{textcomp}%
\usepackage{manyfoot}%
\usepackage{booktabs}%
\usepackage{array}
\usepackage{algorithm}%
\usepackage{algorithmicx}%
\usepackage{algpseudocode}%
\usepackage{listings}%
\usepackage{subfig}
\usepackage{natbib}
\usepackage{hyperref}
\usepackage{pifont}

\usepackage{anyfontsize}
\usepackage{lmodern}
\usepackage{array}
\usepackage{url}
\usepackage{verbatim}
\usepackage{cite}
\usepackage{nicefrac}
\usepackage{geometry}
\usepackage{silence}
\usepackage{colortbl}
\usepackage{subcaption}  
\definecolor{mygray}{gray}{0.95}
\definecolor{mygray2}{gray}{0.75}
\begin{document}

\title[Article Title]{Forensic-Aware Continual Adaptation for Image Forgery Localization}

\author[1]{\fnm{Chenqi} \sur{Kong}}\email{cqkong@nus.edu.sg}


\author[2]{\fnm{Song} \sur{Xia}}\email{xias0002@ntu.edu.sg}

\author[3]{\fnm{Anwei} \sur{Luo}}\email{luoanwei@jxufe.edu.cn}

\author[4]{\fnm{Peisong} \sur{He}}\email{gokeyhps@scu.edu.cn}


\author[2]{\fnm{Alex} \sur{C. Kot}}\email{eackot@ntu.edu.sg}

\author[3]{\fnm{Yuming} \sur{Fang}}\email{fa0001ng@e.ntu.edu.sg}

\affil[1]{\orgdiv{School of Computing}, \orgname{National University of Singapore}, \orgaddress{\country{Singapore}}}

\affil[2]{\orgdiv{ROSE Lab}, \orgname{School of EEE, Nanyang Technological University}, \orgaddress{\country{Singapore}}}

\affil[3]{\orgdiv{School of Computing and Artificial Intelligence}, \orgname{Jiangxi University of Finance and Economics}, \orgaddress{\state{Nanchang}, \country{China}}}

\affil[4]{\orgdiv{School of Cyber
Science and Engineering}, \orgname{Sichuan University}, \orgaddress{\state{Chengdu}, \country{China}}}

\abstract{
The rapid evolution of image manipulation techniques has raised pressing public security concerns.
Existing countermeasures mitigate these threats by accurately localizing manipulated regions in images.
However, they are often unable to adapt dynamically to newly emerging forgeries.
In real-world forensic scenarios, data typically arrive sequentially, yet existing Image Forgery Localization (IFL) methods largely overlook the need for continual model adaptation in practical workflows.
To address this gap, we introduce the first continual learning framework for IFL and establish a comprehensive benchmark under two realistic data-evolution protocols: cross-dataset and cross-content continual learning.
Extensive evaluations of representative state-of-the-art IFL and continual learning methods reveal substantial performance degradation under both settings, exposing two key challenges: (1) how to adaptively capture intrinsic forensic traces from incoming data across diverse unseen domains, and (2) how to preserve previously acquired forensic knowledge during sequential adaptation.
To address these challenges, we propose a forensic-aware continual adaptation framework.
First, we develop a forensic trace mining module with a Spatial Mixture-of-Forensic-Experts (SMoFE) mechanism to dynamically route complementary forensic cues at different spatial locations, together with Forensic Evidence-Guided Dense Prompting (FEGDP) to transform low-level forensic traces into structured localization evidence for SAM.
Second, we introduce a Fisher-weighted LoRA Gradient (FLAG) surgery strategy that identifies old-task-sensitive adaptation directions and selectively suppresses conflicting updates, thereby mitigating catastrophic forgetting while retaining sufficient plasticity for emerging forgery domains.
Extensive experiments demonstrate state-of-the-art performance in both pixel-level forgery localization and image-level forgery detection across diverse continual learning scenarios.}


\maketitle

\section{Introduction}\label{sec1}
Image manipulation has long posed serious security concerns across a wide range of areas, such as politics and fraud.  
To counter these threats, substantial efforts have been made to develop Image Forgery Localization (IFL) techniques to capture various manipulation artifacts, including lens distortions \citep{mayer2018accurate,yerushalmy2011digital}, Color Filter Array (CFA) artifacts \citep{kong2025pixel,cao2009accurate}, noise patterns \citep{lyu2014exposing,kobayashi2010detecting}, compression artifacts \citep{fan2003identification,chen2011detecting}, among others. Beyond these hand-crafted features, later learning-based methods built upon advanced architectures (e.g., SAM \citep{kirillov2023segment} and Swin \citep{liu2021swin}) have achieved higher localization accuracy. Recent studies have further focused on improving model generalization to unseen datasets and robustness against common image distortions.


Despite this progress, existing IFL models still exhibit limited generalizability when encountering newly emerging forgeries. This limitation has become more severe with the advent of frontier foundation models, such as Nano Banana 2 \citep{Nano-Banana2} and GPT-Image 2.5 \citep{GPT-Image2}, which can generate highly realistic content and blur the boundary between authentic and manipulated images. These limitations mainly arise from two challenges. First, newly emerging forgeries are increasingly sophisticated and often leave fewer detectable traces for existing models. Second, new manipulation techniques can be applied to substantially different image domains, such as document and scientific images, resulting in large domain gaps. As foundation models rapidly evolve, IFL defenses often lag behind emerging attacks. Therefore, IFL models must be frequently updated to adapt to new forgeries.
  
\begin{figure}[tp]
  \centering
  \includegraphics[width=0.70\linewidth]{ 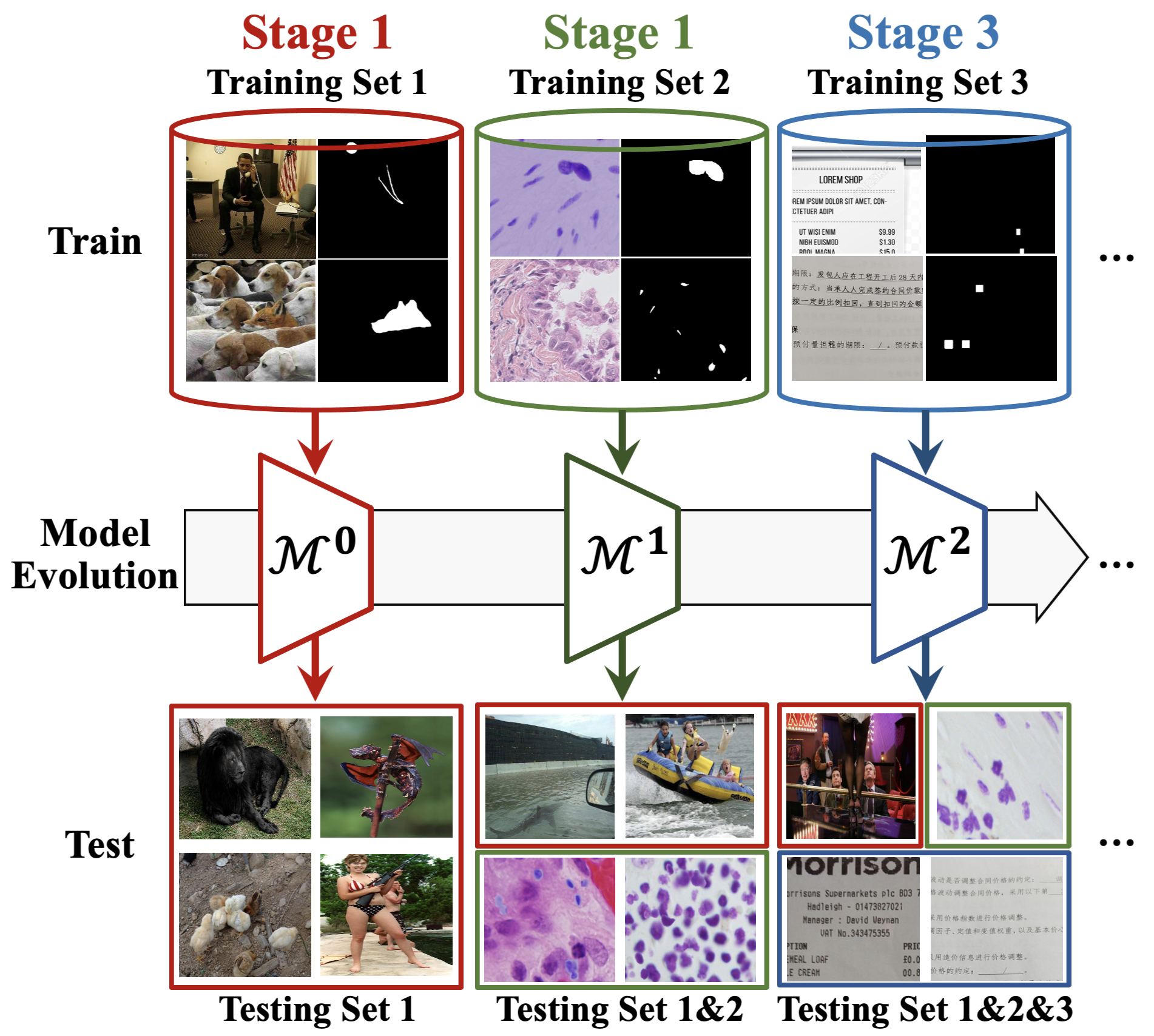}
  \caption{Illustration of Continual Image Forgery Localization (CIFL) pipeline (Best viewed in color). $\mathcal{M}$ indicates different models.}
  \label{fig:teaser}
  \end{figure}
  
However, state-of-the-art IFL models, particularly SAM-based methods \citep{kirillov2023segment, kwon2025safire, peng2025forensicssam}, typically contain a large number of parameters and are trained on datasets with millions of images. Retraining the entire model on previous datasets whenever new forgeries emerge is computationally expensive and impractical for real-world deployment. This motivates the need for a continual IFL paradigm that can progressively adapt IFL models to newly emerging forgeries while retaining previously learned knowledge. As illustrated in Figure~\ref{fig:teaser}, unlike prior studies that formulate IFL as a conventional static learning task, this work recasts IFL as a more practical continual learning problem. Specifically, the model is sequentially updated on a series of newly arriving training sets, resulting in progressive model evolution across different learning stages. After each stage, the updated model is tested on both previously seen and newly introduced test sets, thereby assessing its ability to acquire new forensic knowledge while preserving previously learned capabilities.

To this end, we first conduct an extensive benchmark of state-of-the-art continual learning and IFL methods under two evaluation protocols. \textbf{Protocol 1} consists of four image manipulation datasets with diverse forgery types, namely Classic \citep{Dong2013, IFC, zampoglou2015detecting, novozamsky2020imd2020}, Defacto \citep{mahfoudi2019defacto}, FantasticReality \citep{kniaz2019point}, and TampCOCO \citep{kwon2022learning}, \textbf{sorted according to their release dates}. This protocol simulates practical scenarios in which datasets arrive sequentially. \textbf{Protocol 2} covers three types of image content: natural images \citep{Dong2013, IFC, zampoglou2015detecting, novozamsky2020imd2020}, document images \citep{qu2023towards, yu2023learning}, and scientific images \citep{cardenuto2022benchmarking, sabir2021biofors, lin2024exposing}. This protocol reflects more complex real-world scenarios and introduces more pronounced domain gaps. We benchmarked SOTA continual learning and IFL techniques under both protocols and found that these methods suffer significant performance degradation in the continual learning paradigm.

Based on these observations, we identify two critical challenges in continual IFL:
(1) how to adaptively capture intrinsic forensic traces from incoming data across diverse unseen domains; and
(2) how to preserve previously acquired forensic knowledge while adapting to substantially shifted manipulation distributions.
To address these challenges, we propose a forensic-aware continual adaptation framework that jointly enhances forensic representation learning and knowledge preservation.
For representation adaptation, we introduce Spatial Mixture-of-Forensic-Experts (SMoFE), which performs fine-grained spatial routing over complementary forensic traces, together with Forensic Evidence-Guided Dense Prompting (FEGDP), which bridges the forensic-to-prompt representation gap by transforming low-level artifacts into structured localization evidence for SAM.
For knowledge preservation, we develop a Fisher-weighted LoRA Gradient (FLAG) surgery strategy that constructs an importance-aware adaptation geometry and selectively suppresses updates that conflict with old-task-sensitive directions.
Together, these designs enable the model to continuously acquire emerging forensic patterns while retaining previously learned localization knowledge.

The main contributions of this work are summarized as follows:

\begin{itemize}
\item We systematically study \emph{Continual Image Forgery Localization} (CIFL), reformulating conventional static IFL into a more practical continual learning setting. We further establish cross-dataset and cross-content benchmarks to facilitate future forensic research.

\item We propose a novel forensic-aware representation adaptation framework that goes beyond conventional feature fusion. Specifically, SMoFE performs spatially adaptive Top-1 routing over heterogeneous forensic traces, while FEGDP bridges the forensic-to-prompt representation gap by converting low-level artifacts into structured localization evidence for SAM.

\item We introduce a novel Fisher-guided gradient surgery strategy for knowledge-preserving continual adaptation. Our method establishes an importance-aware geometry and selectively suppresses conflicts along old-task-sensitive directions, thereby improving the stability-plasticity trade-off during continual adaptation.

\item Extensive experiments demonstrate state-of-the-art performance in both pixel-level forgery localization and image-level forgery detection across two continual learning protocols. Ablation experiments further demonstrate the effectiveness and adaptability of the designed modules. 
\end{itemize}

\section{RELATED WORKS}
\subsection{Image forgery localization}
Image forgery localization aims to identify manipulated regions at the pixel level and is a central task in multimedia forensics. Early methods primarily relied on hand-crafted forensic cues, such as sensor noise inconsistencies \citep{lyu2014exposing, kobayashi2010detecting, popescu2004statistical, mahdian2009using, cozzolino2015splicebuster, fan2013estimating}, JPEG compression artifacts \citep{fan2003identification, chen2011detecting, iakovidou2018content, barni2010identification, bianchi2012image, fu2007generalized, pasquini2017statistical}, demosaicing traces \citep{kong2025pixel, cao2009accurate, ferrara2012image, popescu2005exposing, gallagher2008image, ho2010inter}, and lens distortions \citep{mayer2018accurate, yerushalmy2011digital, johnson2006exposing, yerushalmy2011digital, gloe2010efficient, fu2012forgery}, to reveal local tampering evidence. With the rise of deep learning, convolutional neural networks substantially improved localization accuracy by learning forgery-aware representations directly from data. These methods include RGB-based segmentation approaches \citep{lou2025exploring, yao2025dense, chen2025gim}, noise- and residual-guided frameworks \citep{han2024hdf, li2024noise, li2024unionformer}, frequency-domain methods \citep{liu2023explicit, wang2022objectformer, li2019localization, wu2019mantra}, and multi-branch architectures \citep{chen2024ean, sun2026forgerysleuth} that fuse complementary forensic cues. More recently, transformer-based models \citep{lin2023image, kong2025pixel} and large pretrained vision backbones \citep{huang2025sida, xu2025fakeshield, huang2026realign, kwon2025safire} have demonstrated stronger representation capacity and better cross-domain generalization, motivating their adaptation to forgery localization through lightweight fine-tuning and task-specific modules. Nevertheless, robust forgery localization in the wild remains challenging due to weak and heterogeneous manipulation traces and substantial domain gaps across manipulation types and datasets \citep{kong2025pixel}. These challenges indicate that effective localization requires not only strong generic visual priors but also explicit modeling of forensic evidence and adaptation mechanisms that preserve previously acquired manipulation knowledge.

\subsection{Continual learning in forensics}
With the rapid development of AIGC techniques, IFL systems inevitably encounter test samples from unseen domains, making continual adaptation to newly collected data essential. Continual learning methods can be broadly categorized into three groups: (1) regularization-based methods \citep{wang2026learning, kirkpatrick2017overcoming, wang2022learning}, which introduce additional regularization terms to consolidate previously learned knowledge while adapting models to new tasks; (2) replay-based methods \citep{liu2026IDER, kapoor2026hicl}, which selectively store samples from previous tasks in a memory bank and replay them during training on the current task; and (3) parameter-isolation methods \citep{zhang2026grow, ye2026learning}, which typically rely on a task oracle to design task-specific architectures or allocate task-specific parameters. Recent efforts in continual forensic learning have covered a wide range of applications, including face anti-spoofing \citep{cai2023rehearsal, rostami2021detection, perez2020learning}, AIGC detection \citep{tang2025towards, marra2019incremental, wang2022s}, and face forgery detection \citep{cheng2025stacking, zhang2025generalization, tian2024dynamic, sun2025continual}. Although these methods address continual binary classification and alleviate catastrophic forgetting to some extent, continual image forgery localization remains largely unexplored. Unlike binary classification, IFL is a dense prediction task that requires pixel-level localization of manipulated regions, making continual adaptation substantially more challenging. To tackle this problem, we introduce spatially adaptive forensic evidence modeling for fine-grained localization and Fisher-guided gradient regulation for preserving previously acquired knowledge during sequential adaptation.

\section{Methodology}
\subsection{Problem Formulation}
\label{sec:problem_formulation}

Conventional image forgery localization assumes that all training samples are available simultaneously and are drawn from a stationary distribution. In contrast, we consider a more practical \emph{Continual Image Forgery Localization} (CIFL) setting, in which forensic data from different domains arrive sequentially. Formally, let
\begin{equation}
    \mathcal{S}
    =
    \left(
        \mathcal{D}^{[1]},
        \mathcal{D}^{[2]},
        \ldots,
        \mathcal{D}^{[Q]}
    \right)
\end{equation}
denote an ordered stream of $Q$ continual-learning tasks. The training
set of the $q$-th task is defined as
\begin{equation}
    \mathcal{D}^{[q]}
    =
    \left\{
        \left(
            \mathbf{x}^{[q]}_{n},
            \mathbf{m}^{[q]}_{n},
            y^{[q]}_{n}
        \right)
    \right\}_{n=1}^{N_q},
    \qquad q \in \{1,\ldots,Q\},
    \label{eq:task_dataset}
\end{equation}
where $N_q$ is the number of training samples in the current task,
$\mathbf{x}^{[q]}_{n}\in\mathbb{R}^{3\times H\times W}$ denotes an
input image, and
$\mathbf{m}^{[q]}_{n}\in\{0,1\}^{H\times W}$ is its pixel-level manipulation mask. The corresponding image-level forgery label is defined as $y^{[q]}_{n}\in \{0,1\}$.

\subsection{Framework Overview}
\begin{figure}[tp]
  \centering
  \includegraphics[width=1.0\linewidth]{ 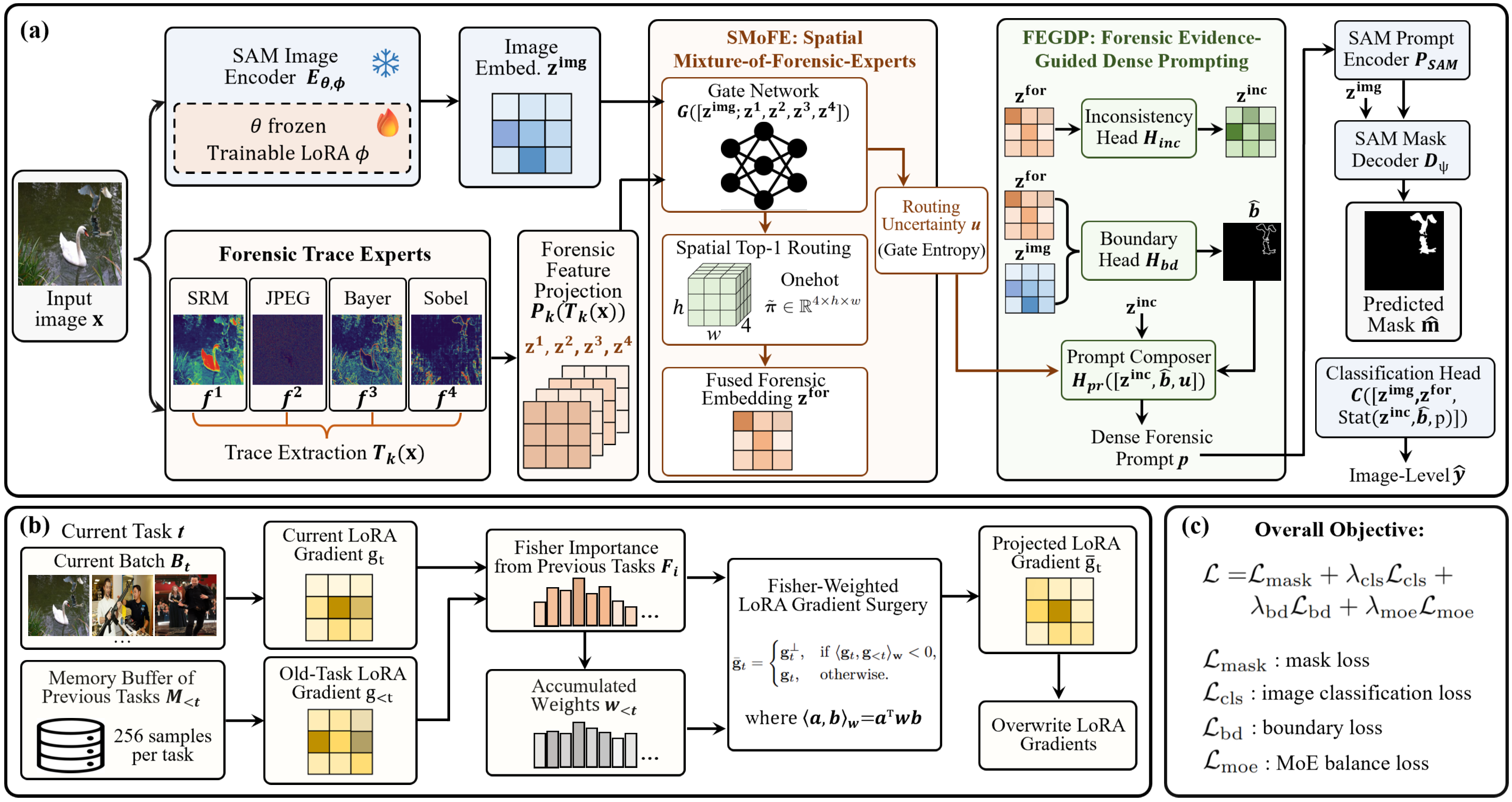}
  \caption{The proposed continual image forgery localization framework. (a) 
  Forensic-Aware Representation Adaptation (FARA) module: Spatial Mixture-of-Forensic-Experts (SMoFE) and Forensic Evidence-Guided Dense Prompting (FEGDP); (b) Fisher-weighted LoRA Gradient (FLAG) surgery module; (c) The overall objectives.}
  \label{fig:framework}
  \end{figure}

Figure~\ref{fig:framework} illustrates the overall framework of the proposed methodology. As shown in Figure~\ref{fig:framework} (a), the model consists of an image encoder $E_{\theta,\phi}$, a forensic prompt learner, and a mask decoder $D_{\psi}$, 
where the pretrained image encoder weights are frozen, while only the LoRA modules in \(E_{\theta,\phi}\) are updated during training. The designed Forensic Trace Mining module consists of a Spatial Mixture-of-Forensic-Experts (SMoFE) module and a Forensic Evidence-Guided Dense Prompting (FEGDP) module. 
SMoFE adaptively combines local anomalies from different forensic maps via MoE mechanisms, and FEGDP aims to transform forensic knowledge into structured prompts that can be readily incorporated into the SAM backbone. 
In addition, we introduce Fisher-guided gradient surgery for continual adaptation, which selectively regulates parameter updates according to their importance to previously learned forensic knowledge, thereby mitigating catastrophic forgetting.

\subsection{Forensic-Aware Representation Adaptation (FARA)}
\label{sec:forensic_prompt_learning}
To adapt forensic representations to evolving manipulation domains, we develop two complementary components. Spatial Mixture-of-Forensic-Experts (SMoFE) adaptively selects informative forensic traces at each spatial location from multiple heterogeneous experts. Based on the routed evidence, Forensic Evidence-Guided Dense Prompting (FEGDP) further transforms low-level forensic responses into structured, localization-oriented prompts that can be effectively consumed by SAM.

\subsubsection{Spatial Mixture-of-Forensic-Experts (SMoFE)}
We build our model upon the Segment Anything Model (SAM) and adapt it to image forgery localization by learning forensic prompts. 
Given an input image $\mathbf{x} \in \mathbb{R}^{3 \times H \times W}$, SAM first extracts visual embeddings
\begin{equation}
    \mathbf{z}^{\mathrm{img}} = E_{\theta,\phi}(\mathbf{x}) \in \mathbb{R}^{C \times h \times w},
\end{equation}
where $E_{\theta,\phi}$ denotes the SAM image encoder equipped with LoRA adapters. 
The pretrained SAM parameters $\theta$ are frozen, while the low-rank adapters $\phi$ are trainable. 

There are two challenges in modeling forensic representations in the continual paradigm: (1) How to expose manipulation artifacts that are not explicitly modeled by natural-image segmentation backbones? (2) How to adaptively extract the most appropriate forensic features of newly incoming data from unseen domains?  

In this work, we introduce a forensic trace branch with multiple artifact experts. Specifically, we use four complementary trace extractors, which have been widely proven effective in previous IFL works:
\begin{equation}
    \mathcal{E} = \{\mathrm{Sobel}, \mathrm{SRM}, \mathrm{JPEG}, \mathrm{Bayer}\},
\end{equation}
These forensic experts respectively capture edge discontinuities \citep{dong2022mvss}, high-frequency residuals \citep{fridrich2012rich}, compression blockiness \citep{fan2003identification}, and CFA/Bayer-pattern \citep{wu2019mantra} inconsistencies. 
For each expert $k \in \mathcal{E}$, a handcrafted forensic trace prior is first extracted and then projected to the SAM embedding space:
\begin{equation}
    \mathbf{z}^{k} = P_k(T_k(\mathbf{x})) \in \mathbb{R}^{C \times h \times w},
\end{equation}
where $T_k(\cdot)$ is the $k$-th forensic trace extractor, and $P_k(\cdot)$ is a learnable convolutional projector. 

Different manipulation types tend to expose forensic artifacts in different local regions. To complementarily integrate these diverse forensic traces, we design a novel spatial Mixture-of-Experts (MoE) mechanism.
Unlike previous MoE works that usually select the optimal expert map along the channel dimension, we perform element-wise MoE in the spatial feature plane. Due to the geometric alignment among the four forensic trace maps, our designed SMoFE can adaptively combine local anomalies from these traces. Such spatially sparse routing allows different locations within the same image to activate different forensic experts, making it particularly suitable for capturing localized and heterogeneous manipulation artifacts.

We design a gate network to predict expert probabilities conditioned on both semantic visual features and forensic trace features:
\begin{equation}
    \boldsymbol{\pi} = 
    \mathrm{softmax}
    \left(
    G\left(
    [\mathbf{z}^{\mathrm{img}}; \mathbf{z}^{1}, \ldots, \mathbf{z}^{K}]
    \right)
    \right),
    \quad
    \boldsymbol{\pi} \in \mathbb{R}^{K \times h \times w}.
\end{equation}
The SAM image embedding $\mathbf{z}^\mathrm{img}$ is  incorporated into the MoE router as contextual guidance rather than as an additional expert. While the forensic experts capture low-level manipulation traces such as edges, noise residuals, compression artifacts, and CFA inconsistencies, the image embedding provides semantic and structural context from the visual backbone. By concatenating image and expert embeddings, the router can make spatially adaptive expert selections conditioned on both forensic evidence and image content. This allows the model to suppress unreliable trace responses in benign regions and emphasize the most relevant forensic cue around manipulated areas. We perform element-wise expert selection as follows:
\begin{equation}
    \tilde{\boldsymbol{\pi}}
    =
    \mathrm{onehot}\left(\arg\max_k \boldsymbol{\pi}_k\right).
\end{equation}
The fused forensic trace embedding is then computed as:
\begin{equation}
    \mathbf{z}^{\mathrm{for}}
    =
    \sum_{k=1}^{K}
    \tilde{\boldsymbol{\pi}}_k \odot \mathbf{z}^{k}.
\end{equation}
In this way, the fused forensic representation integrates complementary local artifacts from different forensic experts. Compared to simply concatenating the four forensic maps, the adopted SMoFE scheme can dynamically select the most relevant forensic expert at each spatial location for newly coming data from unseen domains. By selectively emphasizing the most informative forensic cue at each spatial location, SMoFE also suppresses noisy or domain-irrelevant trace responses, yielding a more flexible forensic representation as the data distribution evolves over time.  

\subsubsection{Forensic Evidence-Guided Dense Prompting (FEGDP)}
The vanilla SAM takes image embeddings together with meaningful dense or sparse prompts (e.g., masks, points, or bounding boxes) as inputs.
Previous SAM-based IFL methods typically inject low-level forensic cues into SAM as prompts.
However, such forensic traces mainly characterize local signal irregularities and are not naturally aligned with the semantic and spatial representations expected by the pretrained SAM prompt-mask decoding pipeline.
This \emph{forensic-to-prompt representation gap} motivates us to transform raw forensic responses into structured manipulation evidence before prompting SAM.
Accordingly, we propose FEGDP to progressively convert heterogeneous forensic traces into localization-oriented evidence, including regional inconsistency, manipulation boundaries, and routing uncertainty, and further compose them into a SAM-compatible dense forensic prompt.

To bridge this representation gap, we decompose forgery evidence into two complementary localization cues: {regional inconsistency}, which identifies suspicious regions exhibiting abnormal forensic statistics, and {boundary evidence}, which captures transition patterns between manipulated and pristine regions.
Such evidence decomposition provides a structured intermediate representation between low-level forensic traces and the spatial prompts consumed by SAM.
We first predict an inconsistency evidence map from the fused forensic embedding:
\begin{equation}
    \mathbf{z}^{\mathrm{inc}}
    =
    H_{\mathrm{inc}}(\mathbf{z}^{\mathrm{for}}),
\end{equation}
where $H_{\mathrm{inc}}(\cdot)$ measures local feature inconsistency through multi-scale neighborhood statistics.
In particular, for each spatial scale $s \in \{3,5,7\}$, we compare the local feature with its neighborhood mean and aggregate cosine dissimilarity and local variance cues.
This encourages the model to highlight regions whose forensic traces are locally inconsistent with their surroundings.
Compared with directly forwarding raw forensic responses, such local-reference modeling emphasizes relative forensic anomalies rather than absolute trace magnitudes, making the resulting evidence less sensitive to domain-dependent variations in forensic statistics.

We further estimate boundary evidence by jointly modeling forensic and visual representations:
\begin{equation}
    \mathbf{\hat{b}}
    =
    H_{\mathrm{bd}}
    \left(
    \mathbf{z}^{\mathrm{for}},
    \mathbf{z}^{\mathrm{img}},
    \mathbf{z}^{\mathrm{inc}}
    \right),
\end{equation}
where $H_{\mathrm{bd}}(\cdot)$ takes the forensic embedding $\mathbf{z}^{\mathrm{for}}$, the SAM image embedding $\mathbf{z}^{\mathrm{img}}$, and the inconsistency evidence $\mathbf{z}^{\mathrm{inc}}$ as inputs.
The forensic embedding provides manipulation-sensitive low-level responses, while the image embedding supplies semantic and structural context.
Their complementary interaction helps distinguish genuine manipulation boundaries from strong but benign image edges, while the inconsistency evidence further provides a spatial prior that focuses boundary estimation on suspicious regions.
In this way, the boundary branch captures fine-grained transition artifacts around manipulated regions.

Beyond deterministic forensic evidence, we further preserve the uncertainty of expert selection rather than discarding it after Top-1 routing.
Specifically, the routing uncertainty is measured by the normalized entropy of the expert probabilities:
\begin{equation}
    \mathbf{u}
    =
    -
    \frac{1}{\log K}
    \sum_{k=1}^{K}
    \boldsymbol{\pi}_k
    \log(\boldsymbol{\pi}_k + \epsilon).
\end{equation}
The normalized gate entropy $\mathbf{u}$ quantifies the uncertainty of the forensic MoE router at each spatial location.
A low entropy indicates that one forensic expert provides a dominant explanation, whereas a high entropy suggests competing or ambiguous evidence among different forensic traces.
Instead of treating such ambiguity as noise, FEGDP explicitly exposes it to the prompt composer, allowing the generated prompt to distinguish confident forensic evidence from regions requiring more cautious evidence integration.

The three signals characterize complementary aspects of manipulation evidence: inconsistency indicates suspicious regions, boundary evidence refines their spatial extent, and routing uncertainty reflects the reliability of the underlying forensic traces.
Rather than directly injecting heterogeneous low-level forensic features into SAM, FEGDP jointly transforms these signals into a localization-oriented dense prompt.
Specifically, the prompt composer is implemented as a lightweight three-layer convolutional network:
\begin{equation}
    \mathbf{p}
    =
    H_{\mathrm{pr}}
    \left(
    [
    \mathbf{z}^{\mathrm{inc}},
    \mathbf{\hat{b}},
    \mathbf{u}
    ]
    \right),
\end{equation}
where $[\cdot]$ denotes concatenation along the channel dimension and $\mathbf{p}$ denotes the resulting dense forensic prompt.
The prompt is subsequently encoded by the SAM prompt encoder and injected into the SAM mask decoder:
\begin{equation}
    \hat{\mathbf{m}}
    =
    D_{\psi}
    \left(
    \mathbf{z}^{\mathrm{img}},
    P_{\mathrm{SAM}}(\mathbf{p})
    \right),
\end{equation}
where $P_{\mathrm{SAM}}$ denotes the SAM prompt encoder and $D_{\psi}$ denotes the SAM mask decoder.

In this manner, FEGDP converts heterogeneous low-level forensic responses into structured, localization-oriented evidence that is better aligned with the pretrained SAM prompting mechanism.
Unlike manually specified point or box prompts, or the direct injection of raw forensic features, the resulting dense prompt explicitly encodes suspicious regions, manipulation boundaries, and forensic uncertainty, providing SAM with task-specific evidence for precise forgery localization.
Importantly, the prompt representation is constructed from generic forensic evidence rather than manipulation-specific semantic categories, making it naturally suitable for continually evolving forgery domains.

\subsection{Fisher-Weighted LoRA Gradient Surgery (FLAG)}
\label{sec:gradient_surgery}





Continual image forgery localization requires adapting to evolving manipulation distributions without overwriting forensic knowledge acquired from previous domains. 
Different domains may exhibit substantially different forensic cues and image contents, thereby leading to catastrophic forgetting. We perform knowledge preservation within the lightweight LoRA adaptation subspace.

A key observation is that different LoRA dimensions contribute unequally to previously learned forensic knowledge.  Treating all parameter dimensions uniformly may restrict less important directions while failing to sufficiently protect those critical to previous tasks. To account for this asymmetry, we propose an importance-aware gradient surgery strategy, which characterizes the importance of individual adaptation directions learned from previous tasks, thereby preserving previously acquired forensic knowledge while retaining sufficient flexibility to learn emerging forgery patterns.


\subsubsection{Fisher-Guided Adaptation Importance Modeling}
Let $\phi$ denote the trainable LoRA parameters of the SAM image encoder.
For each completed task $i<t$, we maintain a small memory buffer $\mathcal{M}_i$ containing 256 representative samples.
During task $t$, a mini-batch from the current task and a reference mini-batch from previous-task memories are used to obtain the corresponding LoRA gradients:
\begin{equation}
    \mathbf{g}_t
    =
    \nabla_{\phi}\mathcal{L}(\mathcal{B}_t;\Theta),
    \qquad
    \mathbf{g}_{<t}
    =
    \nabla_{\phi}\mathcal{L}(\mathcal{B}_{<t};\Theta).
\end{equation}
Here, $\mathbf{g}_t$ represents the adaptation direction preferred by the current forgery domain, whereas $\mathbf{g}_{<t}$ provides a compact estimate of the direction that preserves previously learned forensic knowledge.

A key limitation of conventional gradient-conflict handling is that all parameter dimensions are typically treated uniformly.
However, interference on an old-task-critical LoRA dimension can be considerably more destructive than an equally large conflict on an unimportant dimension.
To explicitly characterize this asymmetry, we estimate the importance of each LoRA parameter using diagonal Fisher information.
After completing task $i$, its Fisher importance is estimated from the stored samples as
\begin{equation}
    \mathbf{F}_i
    =
    \mathbb{E}_{\mathcal{B}\sim\mathcal{M}_i}
    \left[
        \left(
        \nabla_{\phi}
        \mathcal{L}(\mathcal{B};\Theta)
        \right)^2
    \right],
\end{equation}
where the square is computed element-wise, and the resulting importance vectors from all previous tasks are averaged to obtain the accumulated importance weights $\mathbf{w}_{<t}=\frac{1}{t-1}
    \sum_{i=1}^{t-1}
    \mathbf{F}_i$.

In this way, $\mathbf{w}_{<t}$ establishes an {importance-aware geometry} over the LoRA adaptation space:
dimensions that are highly sensitive to previous forensic domains receive larger weights, whereas less relevant dimensions remain comparatively unconstrained.


\subsubsection{Importance-Aware Gradient Surgery}
Given the accumulated Fisher importance $\mathbf{w}_{<t}$, we further use it to regulate gradient interference between the current and previous tasks.
Conventional gradient surgery typically measures gradient conflicts in a uniform Euclidean space, implicitly treating all parameter dimensions as equally important.
However, in continual adaptation, conflicts along directions that are critical to previous tasks should receive greater attention than those occurring along less important directions.
We therefore define a Fisher-weighted inner product as
\begin{equation}
    \langle \mathbf{a},\mathbf{b}\rangle_{\mathbf{w}}
    =
    \sum_j w_{<t}^{j} a^{j} b^{j}.
\end{equation}
A negative value of
$\langle\mathbf{g}_t,\mathbf{g}_{<t}\rangle_{\mathbf{w}}$
indicates that the current update conflicts with previously learned knowledge, particularly along LoRA dimensions that are important to old tasks. When such a conflict occurs, we remove only the component of the current gradient that interferes with the previous-task direction under the Fisher-weighted geometry:

\begin{equation}
    \bar{\mathbf{g}}_t
    =
    \begin{cases}
    \displaystyle
    \mathbf{g}_t
    -
    \frac{
        \langle\mathbf{g}_t,\mathbf{g}_{<t}\rangle_{\mathbf{w}}
    }{
        \|\mathbf{g}_{<t}\|_{\mathbf{w}}^2+\epsilon
    }
    \mathbf{g}_{<t},
    &
    \langle\mathbf{g}_t,\mathbf{g}_{<t}\rangle_{\mathbf{w}}<0,
    \\[10pt]
    \mathbf{g}_t,
    &
    \text{otherwise}.
    \end{cases}
\end{equation}

The projected gradient $\bar{\mathbf{g}}_t$ is used only to update the LoRA parameters, while all remaining trainable components retain their ordinary current-task gradients.
Therefore, previous-task samples define a reference direction for identifying potentially destructive updates. 
Conflicts along old-task-critical LoRA dimensions are emphasized, whereas disagreement along less important dimensions is suppressed.

Consequently, the proposed gradient surgery provides an explicit balance between {stability} and {plasticity}.
It suppresses updates that are likely to overwrite previously acquired forensic cues while preserving sufficient freedom to learn manipulation-specific patterns from newly arriving domains.

\subsection{Overall Objectives}
The overall training objective $\mathcal{L}$ consists of the following four parts: the mask prediction loss $\mathcal{L}_{\mathrm{mask}}$, the image classification loss $\mathcal{L}_{\mathrm{cls}}$, the boundary prediction loss $\mathcal{L}_{\mathrm{bd}}$, and the MoE loss $\mathcal{L}_{\mathrm{moe}}$: 
\begin{equation}
\begin{aligned}
    \mathcal{L}
    =
    &\mathcal{L}_{\mathrm{mask}}
    + \lambda_{\mathrm{cls}} \mathcal{L}_{\mathrm{cls}}
    + \lambda_{\mathrm{bd}} \mathcal{L}_{\mathrm{bd}} 
    + \lambda_{\mathrm{moe}} \mathcal{L}_{\mathrm{moe}},
\end{aligned}
\end{equation}
where $\mathcal{L}_{\mathrm{mask}}$ and $\mathcal{L}_{\mathrm{cls}}$ compute the binary cross-entropy losses between the predicted mask $\mathbf{\hat{m}}$ and class
$\mathbf{\hat{y}}$ and the ground-truth manipulation mask $\mathbf{m}$ and image-level label $y \in \{0,1\}$, respectively. 
We aggregate global visual and forensic features together with evidence-map statistics to predict the image-level manipulation score:
\begin{equation}
    \hat{y}
    =
    C
    \left(
    [
    \mathbf{z}^{\mathrm{img}},
    \mathbf{z}^{\mathrm{for}},
    \mathrm{Stat}(\mathbf{z}^{\mathrm{inc}}, \mathbf{\hat{b}}, \mathbf{p})
    ]
    \right),
\end{equation}
where $\mathrm{Stat}(\cdot)$ includes the mean and top-$k$ mean of each evidence map, and $C$ represents two convolutional layers. The boundary loss supervises the predicted boundary $\mathbf{\hat{b}}$ with the boundary label $\mathbf{b}$:
\begin{equation}
    \mathcal{L}_{\mathrm{bd}}
    =
    \mathrm{Focal}
    \left(
    \mathbf{\hat{b}}, \mathbf{b}
    \right)
    +
    \mathrm{Dice}
    \left(
    \mathbf{\hat{b}}, \mathbf{b}
    \right)
    +
    \left\|
    \mathbf{\hat{b}}
    -
    \partial\hat{\mathbf{m}}
    \right\|_1,
\end{equation}
where the third component makes the boundary evidence consistent with the final SAM prediction, $\partial\hat{\mathbf{m}}$ calculates the boundary of the predicted forgery mask.
Finally, the expert balance loss prevents the Top-1 router from collapsing to a single forensic expert:
\begin{equation}
    \mathcal{L}_{\mathrm{moe}}
    =
    \left\|
    \frac{1}{hw}
    \sum_{i,j}
    \boldsymbol{\pi}_{:,i,j}
    -
    \frac{1}{K}\mathbf{1}
    \right\|_2^2.
\end{equation}

\section{Experiments}
\subsection{Experimental Settings}
\subsubsection{Protocols and Datasets}
\noindent \textbf{Protocol 1} simulates a practical scenario in which the model continually adapts as new datasets emerge. Accordingly, the model under Protocol 1 evolves through the following sequence of tasks: \textbf{Classic} \citep{Dong2013, IFC, zampoglou2015detecting, novozamsky2020imd2020}\textbf{$\rightarrow$DEFACTO}\citep{mahfoudi2019defacto}\textbf{$\rightarrow$FantasticReality} \citep{kniaz2019point}\textbf{$\rightarrow$TampCOCO} \citep{kwon2022learning}, sorted by release date, where Classic incorporates four traditional manipulation datasets: CASIAv2 \citep{Dong2013}, IFC \citep{IFC}, WildWeb \citep{zampoglou2015detecting}, IMD2020 \citep{novozamsky2020imd2020}. These datasets include diverse manipulation types (e.g., splicing, copy-move, and inpainting) with various forgery region areas.
We carefully split the data into training and testing sets to avoid information leakage. The image numbers of the training and testing sets are shown in Table~\ref{tab:dataset_statistics} (a).  

\noindent \textbf{Protocol 2} is tailored to assess IFL models' capability when facing novel forgery content, which encompasses three tasks: Natural (N), Scientific (S), and Document (D) data. We collect natural image data from the following four datasets: CASIAv2 \citep{Dong2013}, IFC \citep{IFC}, WildWeb \citep{zampoglou2015detecting}, 
IMD2020 \citep{novozamsky2020imd2020}. The scientific data consists of three datasets: RSIID \citep{cardenuto2022benchmarking}, 
Biofors \citep{sabir2021biofors}, 
SciSp \citep{lin2024exposing}. For document images, we combine STFD \citep{yu2023learning} with a subset of DocTamper \citep{qu2023towards}. Similarly, we carefully breakdown the training and testing sets to avoid information leakage. 
The image numbers of the training and testing sets are shown in Table~\ref{tab:dataset_statistics} (b). Under Protocol 2, we evaluate all six possible task orders (N$\rightarrow$S$\rightarrow$D, N$\rightarrow$D$\rightarrow$S, D$\rightarrow$S$\rightarrow$N, D$\rightarrow$N$\rightarrow$S, S$\rightarrow$N$\rightarrow$D, and S$\rightarrow$D$\rightarrow$N) in a continual learning setting. Compared with Protocol 1, Protocol 2 introduces substantially larger domain gaps between sequential tasks due to the pronounced content differences among natural, scientific, and document images, thereby providing a more challenging setting for evaluating the continual adaptation capability of IFL models.

\subsubsection{Baseline Methods}
We incorporate eight SOTA baseline methods from continual learning and image forgery localization, primarily published in recent top-tier journals and conferences.

\noindent \textbf{HiDe-Prompt  \citep{wang2023hierarchical}} (NeurIPS'23) boosts prompt-based continual learning under self-supervised pretraining by decomposing the objective into prediction, task identification, and task adaptation and explicitly optimizing them with task-specific prompt ensembles and contrastive regularization.

\noindent \textbf{Eclipse \citep{kim2024eclipse}} (CVPR'24) enables efficient continual panoptic segmentation by freezing the backbone and learning only prompt embeddings, with logit manipulation to reduce drift and forgetting while adapting to new classes.

\begin{table*}[t]
\centering
\caption{Statistics of training and testing datasets of Protocol 1 and Protocol 2.}
\label{tab:dataset_statistics}

\begin{minipage}[t]{0.48\textwidth}
    \centering
    \textbf{(a) Protocol 1}\\[0.5ex]
    \resizebox{\linewidth}{!}{
    \begin{tabular}{lccc}
        \toprule
        Dataset & Sum & Train & Test \\
        \midrule
        Classic & 25,999 & 19,735 & 6,264 \\
        DEFACTO & 27,106 & 21,106 & 6,000 \\
        FantasticReality & 36,000 & 26,677 & 9,323 \\
        TampCOCO & 48,445 & 36,445 & 12,000 \\
        \bottomrule
    \end{tabular}
    }
    
    \vspace{0.5ex}
    \footnotesize Image numbers of the training and testing sets for the four datasets under Protocol 1.
\end{minipage}
\hfill
\begin{minipage}[t]{0.42\textwidth}
    \centering
    \textbf{(b) Protocol 2}\\[0.5ex]
    \resizebox{\linewidth}{!}{
    \begin{tabular}{lccc}
        \toprule
        Dataset & Sum & Train & Test \\
        \midrule
        Natural & 16,946 & 12,582 & 4,364 \\
        Scientific & 8,131  & 5,671  & 2,460 \\
        Document & 16,432 & 12,750 & 3,682 \\
        \bottomrule
    \end{tabular}
    }

    \vspace{0.5ex}
    \footnotesize Image numbers of the training and testing sets for the three datasets under Protocol 2.
\end{minipage}

\end{table*}

\noindent \textbf{Cprompt \citep{gao2024consistent}} (CVPR'24) improves prompt-based continual learning by enforcing training–testing consistency through jointly training all classifiers with prompts and enhancing prompt selection robustness.

\noindent \textbf{SEMA \citep{wang2025self}} (CVPR'25) enables pretrained model-based continual learning to self-expand by reusing or adding modular adapters only when distribution-shift descriptors detect unmet changes, with a router to mix adapter outputs for sub-linear growth and better stability–plasticity.

\noindent \textbf{PIM \citep{kong2025pixel}} (TPAMI'25) designs masked self-attention to capture global pixel inconsistencies, a local dependency stream to mine fine-grained manipulation cues, and a pixel-inconsistency data augmentation strategy to emphasize demosaicing-induced pixel-correlation artifacts.

\begin{table*}[t]
\centering
\caption{Pixel-level continual forgery localization performance on sequential tasks (T1-T4) under Protocol 1. F1 and IoU (fixed threshold of 0.5 for both metrics) scores are reported. }
\label{tab:sam_lora_results}
\resizebox{\textwidth}{!}{
\begin{tabular}{c c cc cc cc cc cc}
\hline
\multirow{2}{*}{Method} & \multirow{2}{*}{Task} 
& \multicolumn{2}{c}{Classic} 
& \multicolumn{2}{c}{DEFACTO} 
& \multicolumn{2}{c}{FantasticReality} 
& \multicolumn{2}{c}{TampCOCO} 
& \multicolumn{2}{c}{Average} \\
& 
& F1 & IoU 
& F1 & IoU 
& F1 & IoU 
& F1 & IoU 
& F1 & IoU \\
\hline					

\multirow{4}{*}{\begin{tabular}{c}
HiDe-Prompt \\
(NeurIPS'23)
\end{tabular}}
& T1 & 0.4955 & 0.4196 & -      & -      & -      & -      & -      & -      & 0.4955 & 0.4196 \\
& T2 & 0.4758 & 0.4085 & 0.1378 & 0.1051 & -      & -      & -      & -      & 0.3068 & 0.2568  \\
& T3 & 0.4661 & 0.4001 & 0.1342 & 0.1023 & 0.1723 & 0.1148 & -      & -      & 0.2575 & 0.2057 \\
& T4 & 0.4614 & 0.3959 & 0.1317 & 0.1007 & 0.2169 & 0.1539 & 0.4156 & 0.3630 & 0.3064 & 0.2534 \\
\hline 

\multirow{4}{*}{\begin{tabular}{c}
Eclipse \\
(CVPR'24)
\end{tabular}}
& T1 & 0.4918 & 0.4363 & -      & -      & -      & -      & -      & -      & 0.4918 & 0.4363 \\
& T2 & 0.4641 & 0.4166 & 0.0935 & 0.0723 & -      & -      & -      & -      & 0.2788 & 0.2444  \\
& T3 & 0.3688 & 0.3232 & 0.0983 & 0.0746 & 0.1858 & 0.1250 & -      & -      & 0.2176 & 0.1743 \\
& T4 & 0.3647 & 0.3196 & 0.0957 & 0.0725 & 0.2116 & 0.1465 & 0.3187 & 0.2702 & 0.2477 & 0.2022 \\
\hline 	

\multirow{4}{*}{\begin{tabular}{c}
Cprompt\\
(CVPR'24)
\end{tabular}}
& T1 & 0.5821 & 0.5363 & -      & -      & -      & -      & -      & -      & 0.5821 & 0.5363 \\
& T2 & 0.0914 & 0.0746 & 0.2767 & 0.2431 & -      & -      & -      & -      & 0.1841 & 0.1588  \\
& T3 & 0.1329 & 0.1071 & 0.0998 & 0.0781 & 0.6593 & 0.5714 & -      & -      & 0.2973 & 0.2522 \\
& T4 & 0.0824 & 0.0680 & 0.1156 & 0.0978 & 0.6188 & 0.5574 & 0.8527 & 0.8253 & 0.4174 & 0.3871 \\
\hline 										

\multirow{4}{*}{\begin{tabular}{c}
SEMA \\
(CVPR'25)
\end{tabular}}
& T1 & 0.4934 & 0.4565 & -      & -      & -      & -      & -      & -      & 0.4934 & 0.4565 \\
& T2 & 0.0614 & 0.0490 & 0.2343 & 0.2044 & -      & -      & -      & -      & 0.1479 & 0.1267  \\
& T3 & 0.0943 & 0.0721 & 0.0813 & 0.0599 & 0.4556 & 0.3617 & -      & -      & 0.2104 & 0.1646 \\
& T4 & 0.0412 & 0.0301 & 0.1067 & 0.0873 & 0.2912 & 0.2384 & 0.6747 & 0.6296 & 0.2784 & 0.2464 \\
\hline 		

\multirow{4}{*}{\begin{tabular}{c}
PIM\\
(TPAMI'25)
\end{tabular}}
& T1 & 0.3836 &  0.3180  & -      & -   & -      & -   & -      & -      & 0.3836 &  0.3180 \\
& T2 & 0.2941 & 0.2190 & 0.1691 & 0.1262 & -      & -      & -      & -      & {0.2316} & {0.1726}  \\ 
& T3 & 0.2559 & 0.1756 & 0.1235 & 0.0814 & 0.4949 & 0.3694 & -      & -      & {0.2914} & {0.2088} \\ 
& T4 & 0.1363 & 0.0913 & 0.0618 & 0.0429 & 0.3702 & 0.2755 & 0.6249 & 0.5491 & {0.2983} & {0.2397} \\
\hline 

\multirow{4}{*}{\begin{tabular}{c}
SAFIRE \\
(AAAI'25)
\end{tabular}}
& T1 & 0.5819 & 0.4979  & -  & -  & - & -  & -  & - & 0.5819 &  0.4979 \\
& T2 & 0.1241 & 0.0794 & 0.1811 & 0.1337 & -      & -  & -  & - & {0.1526} & {0.1065}  \\
& T3 & 0.1865 & 0.1217 & 0.1217 & 0.0821 & 0.3597 & 0.2508 & - & - & {0.2226} & {0.1515} \\
& T4 & 0.1110 & 0.0690 & 0.0773 & 0.0512 & 0.2424 & 0.1534 & 0.4101 & 0.2955 & {0.2102} & {0.1423} \\
\hline 

\multirow{4}{*}{\begin{tabular}{c}
IFA-Net \\
(CVPR'26)
\end{tabular}}
& T1 & 0.5228 & 0.4587 & -      & -      & -      & -      & -      & -      & {0.5228} & {0.4587} \\
& T2 & 0.1190 & 0.0820 & 0.1660 & 0.1304 & -      & -      & -      & -      & {0.1425} & {0.1062}  \\
& T3 & 0.1249 & 0.0833 & 0.0797 & 0.0539 & 0.3776 & 0.2693 & -      & -      & {0.1940} & {0.1355} \\
& T4 & 0.1282 & 0.0867 & 0.0780 & 0.0567 & 0.3816 & 0.2931 & 0.6561 & 0.5775 & {0.3110} & {0.2535} \\
\hline		

\multirow{4}{*}{\begin{tabular}{c}SAM-B\&LoRA \\ (Baseline)\end{tabular}}
& T1 & 0.4606 & 0.4168  & -      & -      & -      & -      & -      & -      & 0.4606 & 0.4168 \\
& T2 & 0.0365 & 0.0292 & 0.2116 & 0.1871 & -      & -      & -      & -      & 0.1241 & 0.1081 \\
& T3 & 0.1695 & 0.1369 & 0.0900 & 0.0901 & 0.6989 & 0.6151 & -      & -      & 0.3195 & 0.2807 \\
& T4 & 0.0841 & 0.0683 & 0.1116 & 0.0947 & 0.5819 & 0.5182 & 0.8197 & 0.7927 & 0.3993 & 0.3685\\
\hline 	


\multirow{4}{*}{\begin{tabular}{c}
FOCAL \\
(Ours)
\end{tabular}}
& T1 & 0.6801 & 0.6310  & -      & -      & -      & -      & -      & -      & \textbf{0.6801} & \textbf{0.6310} \\
& T2 & 0.4813 & 0.4269 & 0.4086 & 0.3583 & -      & -      & -      & -      & \textbf{0.4450} & \textbf{0.3926} \\
& T3 & 0.3884 & 0.3399 & 0.2203 & 0.1811 & 0.6869 & 0.6057 & -      & -      & \textbf{0.4319} & \textbf{0.3756} \\
& T4 & 0.4386 & 0.3923 & 0.1970 & 0.1694 & 0.7761 & 0.7165 & 0.8771 & 0.8512 & \textbf{0.5722} & \textbf{0.5324}\\
\hline 

\end{tabular}
}
\end{table*}

\begin{table*}[t]
\centering
\caption{Image-level continual forgery detection performance on sequential tasks (T1-T4) under Protocol 1. AUC and ACC (threshold: 0.5) scores are reported. }
\label{tab:sam_lora_results_img}
\resizebox{\textwidth}{!}{
\begin{tabular}{c c cc cc cc cc cc}
\hline
\multirow{2}{*}{Method} & \multirow{2}{*}{Task} 
& \multicolumn{2}{c}{Classic} 
& \multicolumn{2}{c}{DEFACTO} 
& \multicolumn{2}{c}{FantasticReality} 
& \multicolumn{2}{c}{TampCOCO} 
& \multicolumn{2}{c}{Average} \\
& 
& AUC & ACC 
& AUC & ACC 
& AUC & ACC 
& AUC & ACC 
& AUC & ACC \\
\hline					
\multirow{4}{*}{\begin{tabular}{c}
HiDe-Prompt\\
(NeurIPS'23)
\end{tabular}}
& T1 & 0.8472 & 0.7696 & -      & -      & -      & -      & -      & -      & 0.8472 & 0.7696 \\
& T2 & 0.7577 & 0.7125 & 0.6052 & 0.5055 & -      & -      & -      & -      & 0.6815 & 0.6090  \\
& T3 & 0.7337 & 0.7024 & 0.6044 & 0.5055 & 0.7931 & 0.5488 & -      & -      & 0.7104 & 0.5856 \\
& T4 & 0.7269 & 0.7035 & 0.6030 & 0.5080 & 0.8122 & 0.5588 & 0.7751 & 0.6805 & 0.7293 & 0.6127 \\
\hline 			

\multirow{4}{*}{\begin{tabular}{c}
Eclipse\\
(CVPR'24)
\end{tabular}}
& T1 & 0.8618 & 0.7720 & -      & -      & -      & -      & -      & -      & 0.8618 & 0.7720 \\
& T2 & 0.7789 & 0.7300 & 0.6425 & 0.5562 & -      & -      & -      & -      & 0.7107 & 0.6431  \\
& T3 & 0.7322 & 0.7066 & 0.6025 & 0.5435 & 0.8840 & 0.6569 & -      & -      & 0.7396 & 0.6357 \\
& T4 & 0.7249 & 0.7048 & 0.6027 & 0.5495 & 0.8932 & 0.6798 & 0.7747 & 0.7120 & 0.7489 & 0.6615 \\
\hline 

\multirow{4}{*}{\begin{tabular}{c}
Cprompt\\
(CVPR'24)
\end{tabular}}
& T1 & 0.9618 & 0.9053 & -      & -      & -      & -      & -      & -      & 0.9618 & \textbf{0.9053} \\
& T2 & 0.5272 & 0.5378 & 0.7998 & 0.6893 & -      & -      & -      & -      & 0.6635 & 0.6136  \\
& T3 & 0.4168 & 0.5741 & 0.6493 & 0.6070 & 0.9998 & 0.9942 & -      & -      & 0.6886 & 0.7251 \\
& T4 & 0.5124 & 0.4015 & 0.6678 & 0.5728 & 0.9835 & 0.8943 & 0.9798 & 0.9237 & 0.7859 & 0.6981 \\
\hline 

\multirow{4}{*}{\begin{tabular}{c}
SEMA \\
(CVPR'25)
\end{tabular}}
& T1 & 0.9317 & 0.8736 & -      & -      & -      & -      & -      & -      & 0.9317 & 0.8736 \\
& T2 & 0.5145 & 0.5209 & 0.7657 & 0.6650 & -      & -      & -      & -      & 0.6401 & 0.5930  \\
& T3 & 0.6185 & 0.6427 & 0.5821 & 0.5568 & 0.9918 & 0.9571 & -      & -      & 0.7308 & 0.7189 \\
& T4 & 0.6179 & 0.4566 & 0.6508 & 0.5598 & 0.9577 & 0.7517 & 0.9288 & 0.8028 & 0.7888 & 0.6427 \\
\hline			 				

\multirow{4}{*}{\begin{tabular}{c}
PIM\\
(TPAMI'25)
\end{tabular}}
& T1 & 0.8420 & 0.7604  & -  & - & -  & - & -     & -  & 0.8420 & 0.7604\\ 
& T2 & 0.7906 & 0.7200 & 0.6184 & 0.5102 & -      & - & - & - & {0.7045} & {0.6151}  \\ 
& T3 & 0.3668 & 0.6965 & 0.4974 & 0.5000 & 0.9966 & 0.5199 & - & - & {0.6203} & {0.5721} \\
& T4 & 0.5902 & 0.6967 & 0.5340 & 0.5000 & 0.9237 & 0.5199 & 0.8827 & 0.6783 & {0.7327} & {0.5987} \\
\hline 

\multirow{4}{*}{\begin{tabular}{c}
SAFIRE\\
(AAAI'25)
\end{tabular}}
& T1 & 0.8786 & 0.8027 & -  & -   & -  & -      & -  & -  & 0.8786 & 0.8027 \\
& T2 & 0.6680 & 0.7184 & 0.6664 & 0.5158 & -      & -      & -      & -      & {0.6672} & {0.6171}  \\
& T3 & 0.7413 & 0.7000 & 0.5076 & 0.4993 & 0.9674 & 0.5284 & -      & -      & {0.7388} & {0.5759} \\ 
& T4 & 0.6568 & 0.6967 & 0.5203 & 0.5000 & 0.8359 & 0.5203 & 0.8412 & 0.6783 & {0.7136} & {0.5988} \\ 
\hline 

\multirow{4}{*}{\begin{tabular}{c}
IFA-Net\\
(CVPR'26)
\end{tabular}}
& T1 & 0.8706 & 0.7917 & -      & -      & -      & -      & -      & -      & {0.8706} & {0.7917} \\
& T2 & 0.5715 & 0.6729 & 0.6901 & 0.5392 & -      & -      & -      & -      & {0.6308} & {0.6060}  \\
& T3 & 0.6705 & 0.7087 & 0.5417 & 0.5063 & 0.9782 & 0.6465 & -      & -      & {0.7301} & {0.6205} \\
& T4 & 0.5681 & 0.6983 & 0.5770 & 0.5188 & 0.9110 & 0.6101 & 0.9094 & 0.6852 & {0.7414} & {0.6281} \\
\hline 			
 
\multirow{4}{*}{\begin{tabular}{c}SAM-B\&LoRA \\ (Baseline)\end{tabular}}
& T1 & 0.8966 & 0.8151  & -      & -      & -      & -      & -      & -      & 0.8966 & 0.8151 \\
& T2 & 0.5891 & 0.5350 & 0.7716 & 0.6787 & -      & -      & -      & -      & 0.6803 & 0.6068 \\
& T3 & 0.5238 & 0.6062 & 0.6185 & 0.5722 & 0.9990 & 0.9815 & -      & -      & 0.7138 & 0.7199 \\
& T4 & 0.4905 & 0.4255 & 0.6633 & 0.5733 & 0.9884 & 0.9189 & 0.9714 & 0.8972 & 0.7784 & 0.7037\\
\hline 		

\multirow{4}{*}{\begin{tabular}{c}
FOCAL\\
(Ours)
\end{tabular}}
& T1 & 0.9646 & 0.9029 & -      & -      & -      & -      & -      & -      & \textbf{0.9646} & {0.9029} \\ 
& T2 & 0.8423 & 0.7816 & 0.8343 & 0.6970 & -      & -      & -      & -      & \textbf{0.8383} & \textbf{0.7393}  \\ 
& T3 & 0.7951 & 0.7459 & 0.6604 & 0.5898 & 0.9996 & 0.9899 & - & - & \textbf{0.8183} & \textbf{0.7752} \\ 
& T4 & 0.8284 & 0.7106 & 0.7106 & 0.6120 & 0.9931 & 0.9467 & 0.9822 & 0.9347 & \textbf{0.8786} & \textbf{0.8010} \\ 
\hline

\end{tabular}
}
\end{table*}

\noindent \textbf{SAFIRE \citep{kwon2025safire}} (AAAI'25)
localizes forgeries by using point prompts to segment and partition an image into multiple source regions (rather than binary masks), enabling more stable learning of within-source consistency and improving both multi-source and binary forgery localization.

\noindent \textbf{IFA-Net \citep{zhang2026detecting}} (CVPR'26) localizes manipulations by using a frozen masked autoencoder (MAE) as a realness prior in a closed-loop scheme that fuses reconstruction residuals for coarse masks and then injects task-adaptive prompts to amplify MAE reconstruction failures for refined localization.

\noindent \textbf{SAM-B\&LoRA \citep{kirillov2023segment}.} We use SAM-Base with LoRA as the baseline model. Only LoRA modules are updated while other parameters in the image encoder are frozen during training. 

\subsubsection{Evaluation Metrics}
For image-level forgery detection, we use accuracy (ACC) and the Area Under the Receiver Operating Characteristic Curve (AUC) as the primary metrics. AUC measures the area under the Receiver Operating Characteristic (ROC) curve. For pixel-level forgery localization, following previous methods \citep{kong2025pixel, dong2022mvss, kwon2022learning}, we use F1 and IoU scores as primary metrics.
F1 Score calculates the harmonic mean of precision and recall:   
\begin{equation}
    \label{F1}
 F1 = 2 \times \frac{Precision \times Recall} {Precision + Recall} = \frac{2 \times TP} {2 \times TP + FP + FN},
\end{equation}
where $TP$, $TN$, $FP$, and $FN$ are True Positives, True Negatives, False Positives, and False Negatives.  
The numerator of the IoU metric measures the area of intersection between prediction $P$ and ground-truth $G$, while the denominator calculates the area of the union between $P$ and $G$: 
\begin{equation}
    \label{IoU}
     IoU = \frac{P \cap G}{P \cup G}.
\end{equation}

\subsubsection{Implementation Details} 

Our model is implemented in PyTorch \citep{paszke2019pytorch} and trained on one NVIDIA RTX PRO 6000 GPU. We adopt the SAM ViT-B backbone initialized from the pretrained SAM checkpoint, and insert LoRA modules with rank 32 into the image encoder. The input image size is 512 $\times$ 512. We use the Adam optimizer \citep{kingma2014adam} with $\beta_{1}$=0.9 and $\beta_{2}$=0.999 to train the model with batch size 64. The learning rate and weight decay are 1e-4 and 1e-5, respectively. The model is trained for 10 epochs for each continual-learning task and validated after every epoch on all seen validation sets.


\subsection{Experimental Results under Protocol 1}
The results in Tables~\ref{tab:sam_lora_results} and~\ref{tab:sam_lora_results_img} demonstrate the superiority of the proposed FOCAL framework in continual forgery detection at both pixel and image levels. Compared with existing continual learning methods and IFL baselines, FOCAL consistently achieves the best average performance across the sequential tasks. At the pixel level, FOCAL achieves the highest average F1 and IoU, indicating its stronger ability to preserve localization knowledge while adapting to newly arriving forgery domains. At the image level, FOCAL also achieves the best average AUC and ACC, showing that the learned forensic representations are effective not only for dense localization but also for global forgery discrimination.
The consistent improvements across pixel-level localization and image-level discrimination suggest that the learned forensic representations preserve useful manipulation evidence at both local and global scales during sequential adaptation.
These gains are consistent with the joint design of FOCAL, where spatially adaptive forensic modeling continuously incorporates informative traces from newly arriving domains, while importance-aware gradient regulation prevents such adaptation from excessively overwriting previously acquired knowledge. These results verify that FOCAL provides an effective continual learning scheme for forgery detection by jointly enhancing forensic feature mining and knowledge preservation.

\begin{figure}[tp]
  \centering
  \includegraphics[width=1.0\linewidth]{ 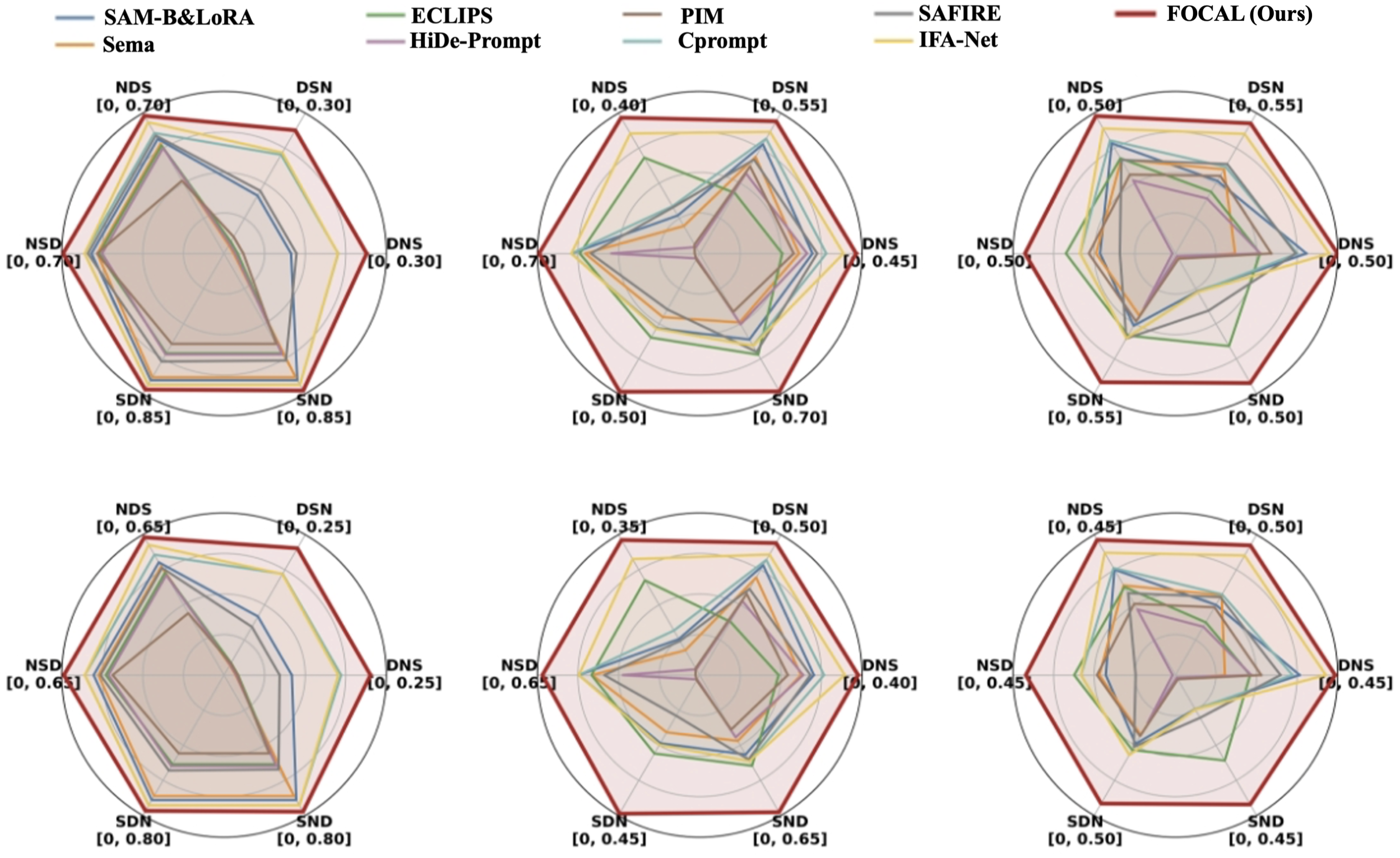}
  \caption{Experimental results under Protocol 2. The top and bottom rows of radar charts show the average F1 and IoU scores, respectively, for different methods across T1-T3 from left to right.}
  \label{fig:Radar}
  \end{figure}

\begin{figure*}[tp]
  \centering
  \includegraphics[width=1.0\linewidth]{ 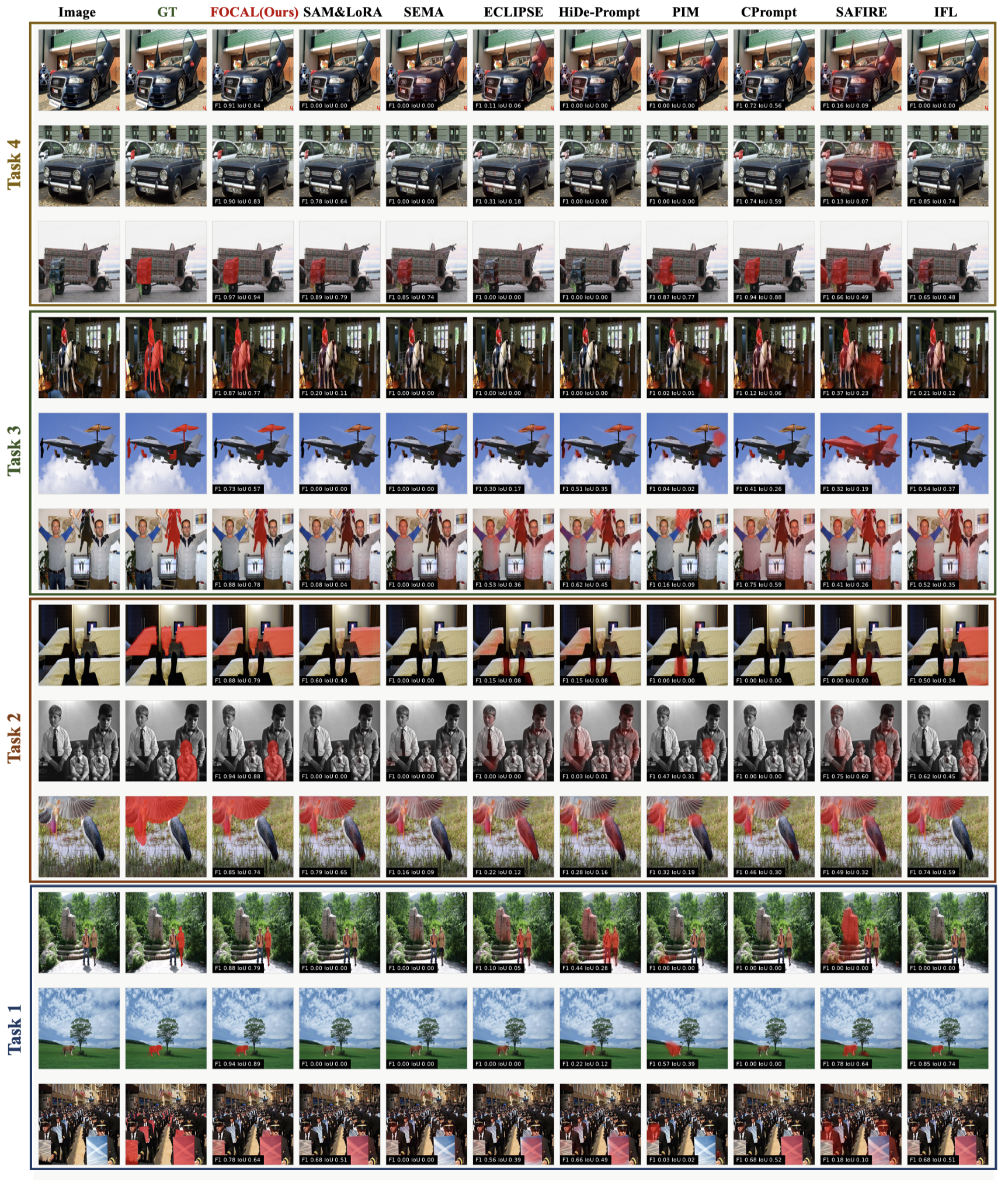}
  \caption{Forgery localization results across Task 1  to Task 4. Models are trained under Protocol 1. The three leftmost columns show the input images, corresponding ground-truth masks, and localization results of our method. The eight rightmost columns present the results of the baseline methods.}
  \label{fig:vis1}
  \end{figure*}
  
\subsection{Experimental Results under Protocol 2}
Figure~\ref{fig:Radar} shows the experimental results under Protocol 2. Each vertex of a radar chart represents a continual setting (e.g., NDS: Natural$\rightarrow$Document$\rightarrow$Scientific). 
The top and bottom rows of radar
charts show the average F1 and IoU scores, respectively, for different methods across
T1-T3 from left to right. For instance, the middle column presents the average F1 and IoU scores across the first two datasets after the model trained on Task 2.
The proposed FOCAL achieves the largest radar coverage across all six task orders under Protocol 2, indicating its strong robustness when facing novel image content from natural, scientific, and document domains. Compared with existing IFL methods, FOCAL consistently achieves promising performance on different order permutations, while the competing methods show obvious fluctuations across task orders. This suggests that FOCAL is less sensitive to the order of incoming domains and can better preserve previously learned forgery knowledge while adapting to new content types. In contrast, methods such as SAM-B\&LoRA, HiDe-Prompt, and PIM suffer from unstable performance under certain orders, revealing their limited ability to handle large domain shifts. IFA-Net and SAFIRE provide relatively competitive results, but they still fall behind FOCAL in terms of overall stability and coverage. These results demonstrate the effectiveness of FOCAL in content-level incremental forgery localization, especially under challenging cross-domain continual learning scenarios.

\subsection{Continual Forgery Localization Results}
The qualitative comparisons in Figure~\ref{fig:vis1} further demonstrate the effectiveness of \textbf{FOCAL} for continual forgery localization. The models are sequentially trained from Task 1 to Task 4, with all results evaluated after completing Task 4. This setting jointly assesses the ability to adapt to newly introduced forgery patterns and retain localization knowledge acquired from previous tasks. FOCAL consistently produces accurate and spatially compact localization masks across all four tasks. For the latest Task 4, FOCAL effectively adapts to newly introduced forgery patterns, while maintaining outstanding localization performance on the previously learned Tasks 1-3. Even after sequential adaptation through all four tasks, FOCAL accurately identifies manipulated regions across diverse scenes involving small objects, irregular regions, human subjects, and complex foreground structures. In contrast, the baseline methods frequently suffer from under- or over-localization and even complete localization failures, showing missing masks and extensive spurious activations. Moreover, some baselines' predictions are considerably less consistent across different tasks, indicating catastrophic forgetting issues and limited adaptability to evolving forgery patterns. Overall, these qualitative results demonstrate that FOCAL achieves a favorable {stability-plasticity trade-off}, effectively adapting to newly emerging forgery patterns while preserving previously acquired forensic knowledge, thereby enabling robust and consistent forgery localization throughout the continual learning process.

\begin{figure*}[tp]
  \centering
  \includegraphics[width=1.0\linewidth]{ 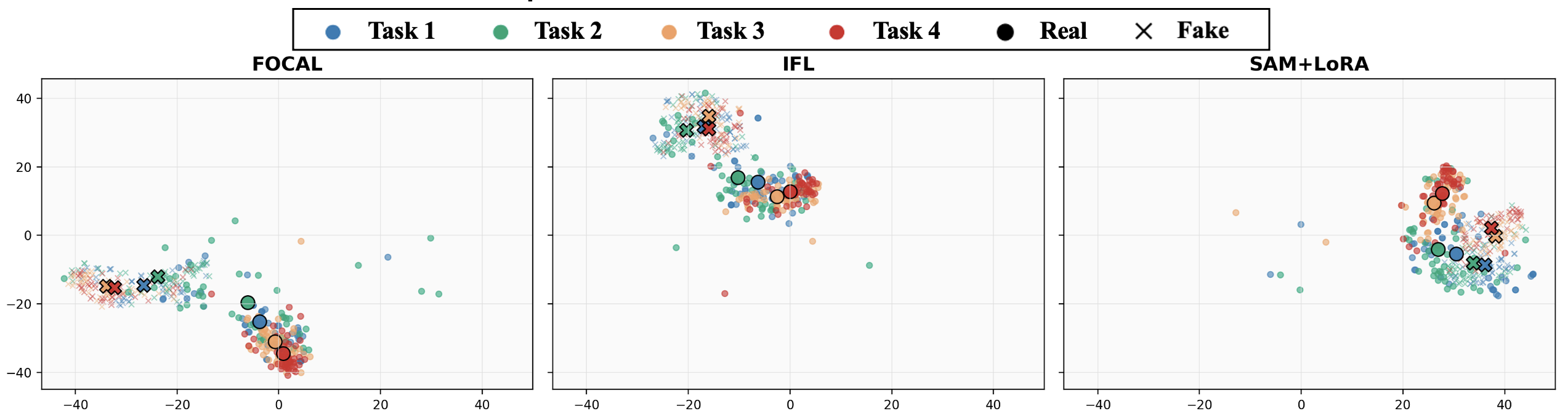}
  \caption{T-SNE distributions from Task 1 to Task 4 under Protocol 1 for our proposed FOCAL, the SOTA method IFL, and the baseline SAM+LoRA.}
  \label{fig:tsne1}
  \end{figure*}
\begin{figure*}[tp]
  \centering
  \includegraphics[width=1.0\linewidth]{ 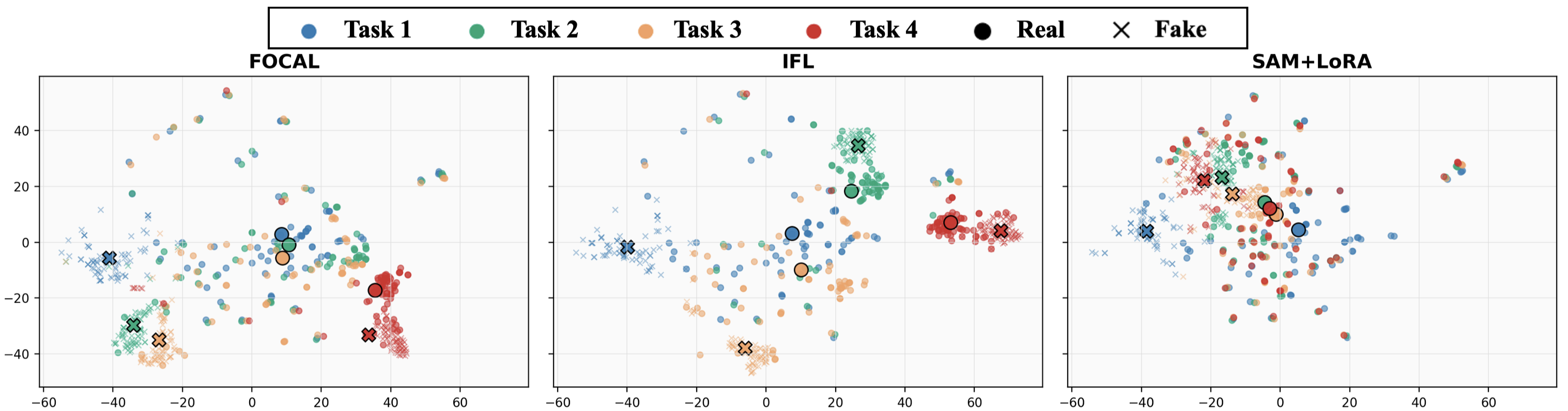}
  \caption{T-SNE visualizations of the Task 1 test data after sequential training on Tasks 1–4, comparing our FOCAL method with the SOTA method IFL and the baseline SAM+LoRA.}
  \label{fig:tsne2}
  \end{figure*}

\begin{table*}[t]
\centering
\caption{Effectiveness of FOCAL on different backbones. }
\label{tab:abl_backbone}
\resizebox{\textwidth}{!}{
\begin{tabular}{c c cc cc cc cc cc}
\hline
\multirow{2}{*}{Setting} & \multirow{2}{*}{Task} 
& \multicolumn{2}{c}{Classic} 
& \multicolumn{2}{c}{DEFACTO} 
& \multicolumn{2}{c}{FantasticReality} 
& \multicolumn{2}{c}{TampCOCO} 
& \multicolumn{2}{c}{Average} \\
& 
& F1 & IoU 
& F1 & IoU 
& F1 & IoU 
& F1 & IoU 
& F1 & IoU \\
\hline					

\multirow{4}{*}{\begin{tabular}{c}SAM-Tiny \end{tabular}}
& T1 & 0.5369 & 0.4982  & -      & -      & -      & -      & -      & -      & 0.5369 & 0.4982 \\ 
& T2 & 0.0615 & 0.0500 & 0.2055 & 0.1808 & -      & -      & -      & -      & 0.1335 & 0.1154 \\ 
& T3 & 0.1545 & 0.1265 & 0.0859 & 0.0678 & 0.6555 & 0.5748 & -      & -      & 0.2986 & 0.2564 \\
& T4 & 0.0947 & 0.0777 & 0.0739 & 0.0621 & 0.5075 & 0.4465 & 0.8045 & 0.7737 & 0.3702 & 0.3400\\
\hline 

\multirow{4}{*}{\begin{tabular}{c}SAM-Tiny\\ + FOCAL \end{tabular}}
& T1 & 0.5571 & 0.5152  & -      & -      & -      & -      & -      & -      & \textbf{0.5571} & \textbf{0.5152} \\
& T2 & 0.1562 & 0.1328 & 0.2190 & 0.1937 & -      & -      & -      & -      & \textbf{0.1876} & \textbf{0.1633} \\ 
& T3 & 0.2220 & 0.1826 & 0.0768 & 0.0595 & 0.6687 & 0.5902 & -      & -      & \textbf{0.3225} & \textbf{0.2774} \\ 
& T4 & 0.1404 & 0.1195 & 0.1050 & 0.0878 & 0.5766 & 0.5126 & 0.8293 & 0.8013 & \textbf{0.4128} & \textbf{0.3803}\\
\hline 
\hline 
\multirow{4}{*}{\begin{tabular}{c}SAM-Base \end{tabular}}
& T1 & 0.4606 & 0.4168  & -      & -      & -      & -      & -      & -      & 0.4606 & 0.4168 \\
& T2 & 0.0365 & 0.0292 & 0.2116 & 0.1871 & -      & -      & -      & -      & 0.1241 & 0.1081 \\
& T3 & 0.1695 & 0.1369 & 0.0900 & 0.0901 & 0.6989 & 0.6151 & -      & -      & 0.3195 & 0.2807 \\
& T4 & 0.0841 & 0.0683 & 0.1116 & 0.0947 & 0.5819 & 0.5182 & 0.8197 & 0.7927 & 0.3993 & 0.3685\\
\hline 

\multirow{4}{*}{\begin{tabular}{c}
SAM-Base\\ + FOCAL \\
\end{tabular}}
& T1 & 0.6801 & 0.6310  & -      & -      & -      & -      & -      & -      & \textbf{0.6801} & \textbf{0.6310} \\
& T2 & 0.4813 & 0.4269 & 0.4086 & 0.3583 & -      & -      & -      & -      & \textbf{0.4450} & \textbf{0.3926} \\
& T3 & 0.3884 & 0.3399 & 0.2203 & 0.1811 & 0.6869 & 0.6057 & -      & -      & \textbf{0.4319} & \textbf{0.3756} \\
& T4 & 0.4386 & 0.3923 & 0.1970 & 0.1694 & 0.7761 & 0.7165 & 0.8771 & 0.8512 & \textbf{0.5722} & \textbf{0.5324}\\
\hline 
\hline 
\multirow{4}{*}{\begin{tabular}{c}SAM-Large\end{tabular}}
& T1 & 0.7696 & 0.7279  & -      & -      & -      & -      & -      & -      & 0.7696 & 0.7279 \\ 
& T2 & 0.2130 & 0.1824 & 0.5573 & 0.5024 & -      & -      & -      & -      & 0.3851 & 0.3424 \\ 
& T3 & 0.2431 & 0.2116 & 0.2366 & 0.1977 & 0.7945 & 0.7330 & -      & -      & 0.4247 & 0.3807 \\  
& T4 & 0.1507 & 0.1348 & 0.2289 & 0.1952 & 0.7257 & 0.6661 & 0.9208 & 0.9015 & 0.5065 & 0.4744\\
\hline 

\multirow{4}{*}{\begin{tabular}{c}SAM-Large\\ + FOCAL \end{tabular}}
& T1 & 0.8050 & 0.7630  & -      & -      & -      & -      & -      & -      & \textbf{0.8050} & \textbf{0.7630} \\ 
& T2 & 0.4962 & 0.4497 & 0.6064 & 0.5495 & -      & -      & -      & -      & \textbf{0.5513} & \textbf{0.4996} \\
& T3 & 0.4930 & 0.4498 & 0.3867 & 0.3299 & 0.8235 & 0.7645 & -      & -      & \textbf{0.5677} & \textbf{0.5147} \\ 
& T4 & 0.4458 & 0.4025 & 0.3216 & 0.2753 & 0.7868 & 0.7306 & 0.9119 & 0.8919 & \textbf{0.6165} & \textbf{0.5751}\\
\hline

\end{tabular}
}
\end{table*}

\begin{table*}[t]
\centering
\caption{Effectiveness of the designed modules. }
\label{tab:module}
\resizebox{\textwidth}{!}{
\begin{tabular}{c c cc cc cc cc cc}
\hline
\multirow{2}{*}{Setting} & \multirow{2}{*}{Task} 
& \multicolumn{2}{c}{Classic} 
& \multicolumn{2}{c}{DEFACTO} 
& \multicolumn{2}{c}{FantasticReality} 
& \multicolumn{2}{c}{TampCOCO} 
& \multicolumn{2}{c}{Average} \\
& 
& F1 & IoU 
& F1 & IoU 
& F1 & IoU 
& F1 & IoU 
& F1 & IoU \\
\hline					

\multirow{4}{*}{\begin{tabular}{c}
FOCAL \\
\end{tabular}}
& T1 & 0.6801 & 0.6310  & -      & -      & -      & -      & -      & -      & \textbf{0.6801} & \textbf{0.6310} \\
& T2 & 0.4813 & 0.4269 & 0.4086 & 0.3583 & -      & -      & -      & -      & \textbf{0.4450} & \textbf{0.3926} \\
& T3 & 0.3884 & 0.3399 & 0.2203 & 0.1811 & 0.6869 & 0.6057 & -      & -      & \textbf{0.4319} & \textbf{0.3756} \\
& T4 & 0.4386 & 0.3923 & 0.1970 & 0.1694 & 0.7761 & 0.7165 & 0.8771 & 0.8512 & \textbf{0.5722} & \textbf{0.5324}\\
\hline 

\multirow{4}{*}{\begin{tabular}{c}w/o FARA\end{tabular}}
& T1 & 0.6721 & 0.6256  & -      & -      & -      & -      & -      & -      & {0.6721} & {0.6256}\\ 
& T2 & 0.1841 & 0.1560 & 0.3982 & 0.3525 & -      & -      & -      & -      & {0.2911} & {0.2543} \\ 
& T3 & 0.1828 & 0.1501 & 0.1452 & 0.1174 & 0.7142 & 0.6340 & -      & -      & {0.3474} & {0.3005} \\
& T4 & 0.1164 & 0.0992 & 0.1471 & 0.1252 & 0.6419 & 0.5788 & 0.8490 & 0.8245 & 0.4386 & 0.4069\\
\hline 

\multirow{4}{*}{\begin{tabular}{c}w/o FLAG\end{tabular}}
& T1 & {0.6801} & 0.6310  & -      & -      & -      & -      & -      & -      & {0.6801} & 0.6310 \\  
& T2 & 0.4463 & 0.3937 & 0.3670 & 0.3221 & -      & -      & -      & -      & {0.4067} & {0.3579} \\
& T3 & 0.3614 & 0.3047 & 0.2125 & 0.1756 & 0.6615 & 0.5918 & -  & -  & {0.4118} & {0.3574} \\ 
& T4 & 0.3955 & 0.3559 & 0.1637 & 0.1398 & 0.7202 & 0.6585 & 0.8796 & 0.8574 & {0.5398} & {0.5029}\\
\hline 
 
\end{tabular}
}
\end{table*}


\subsection{T-SNE Distributions}
Figure~\ref{fig:tsne1} presents the t-SNE feature distributions obtained after sequential training from Task~1 to Task~4 under Protocol~1. FOCAL exhibits a more consistent separation between real and fake samples across the four tasks, with the advantage becoming more evident on later tasks where inter-task interference is stronger. In comparison, SAM+LoRA produces more entangled real/fake distributions despite forming relatively compact clusters, indicating limited discriminative preservation under continual adaptation. IFL improves upon the plain baseline, but FOCAL maintains clearer class separation on most tasks, particularly Tasks~3 and~4, where it achieves substantially higher silhouette scores. These results indicate that FOCAL better preserves discriminative feature structures while adapting to newly introduced forgery domains, thereby improving continual forgery localization.

\begin{table*}[t]
\centering
\caption{Effectiveness of the used forensic experts. }
\label{tab:abl_moe}
\resizebox{\textwidth}{!}{
\begin{tabular}{c c cc cc cc cc cc}
\hline
\multirow{2}{*}{Setting} & \multirow{2}{*}{Task} 
& \multicolumn{2}{c}{Classic} 
& \multicolumn{2}{c}{DEFACTO} 
& \multicolumn{2}{c}{FantasticReality} 
& \multicolumn{2}{c}{TampCOCO} 
& \multicolumn{2}{c}{Average} \\
& 
& F1 & IoU 
& F1 & IoU 
& F1 & IoU 
& F1 & IoU 
& F1 & IoU \\
\hline					

\multirow{4}{*}{\begin{tabular}{c}
FOCAL \\
\end{tabular}}
& T1 & 0.6801 & 0.6310  & -      & -      & -      & -      & -      & -      & \textbf{0.6801} & \textbf{0.6310} \\
& T2 & 0.4813 & 0.4269 & 0.4086 & 0.3583 & -      & -      & -      & -      & \textbf{0.4450} & \textbf{0.3926} \\
& T3 & 0.3884 & 0.3399 & 0.2203 & 0.1811 & 0.6869 & 0.6057 & -      & -      & \textbf{0.4319} & \textbf{0.3756} \\
& T4 & 0.4386 & 0.3923 & 0.1970 & 0.1694 & 0.7761 & 0.7165 & 0.8771 & 0.8512 & \textbf{0.5722} & \textbf{0.5324}\\
\hline 

\multirow{4}{*}{\begin{tabular}{c}w/o Sobel\end{tabular}}
& T1 & 0.6571 & 0.6096  & -      & -      & -      & -      & -      & -      & 0.6571 & 0.6096\\  
& T2 & 0.3878 & 0.3435 & 0.3957 & 0.3499 & -      & -      & -      & -      & {0.3917} & {0.3467} \\
& T3 & 0.3756 & 0.3357 & 0.2108 & 0.1809 & 0.6789 & 0.6001 & -      & -      & {0.4218} & {0.3722} \\ 
& T4 & 0.3739 & 0.3369 & 0.1476 & 0.1272 & 0.7092 & 0.6540 & 0.8762 & 0.8548 & 0.5267 & 0.4932\\
\hline 

\multirow{4}{*}{\begin{tabular}{c}w/o SRM\end{tabular}}
& T1 & 0.6753 & 0.6253  & -      & -      & -      & -      & -      & -      & 0.6753 & 0.6253 \\ 
& T2 & 0.3748 & 0.3307 & 0.3987 & 0.3509 & -      & -      & -      & -      & 0.3867 & 0.3408 \\ 
& T3 & 0.3797 & 0.3381 & 0.2161 & 0.1791 & 0.6408 & 0.5609 & -      & -      & 0.4122 & 0.3594 \\
& T4 & 0.3924 & 0.3545 & 0.1725 & 0.1492 & 0.6289 & 0.5695 & 0.8636 & 0.8402 & 0.5143 & 0.4783\\
\hline 

\multirow{4}{*}{\begin{tabular}{c}w/o JPEG\end{tabular}}
& T1 & 0.6634 & 0.6126  & -      & -      & -      & -      & -      & -      & {0.6634} & {0.6126} \\ 
& T2 & 0.3476 & 0.3038 & 0.4149 & 0.3632 & -      & -      & -      & -      & 0.3812 & 0.3335 \\ 
& T3 & 0.3576 & 0.3287 & 0.2282 & 0.1903 & 0.6800 & 0.6029 & -      & -      & {0.4219} & {0.3740} \\ 
& T4 & 0.3804 & 0.3427 & 0.1504 & 0.1289 & 0.7246 & 0.6663 & 0.8736 & 0.8523 & {0.5323} & {0.4975}\\
\hline 

\multirow{4}{*}{\begin{tabular}{c}w/o Bayer\end{tabular}}
& T1 & 0.6713 & 0.6225  & -      & -      & -      & -      & -      & -      & {0.6713} & {0.6225} \\ 
& T2 & 0.3767 & 0.3321 & 0.3902 & 0.3453 & -      & -      & -      & -      & 0.3835 & 0.3387 \\ 
& T3 & 0.3059 & 0.2665 & 0.1869 & 0.1542 & 0.5907 & 0.5116 & -      & -      & 0.3612 & 0.3107 \\
& T4 & 0.3240 & 0.2894 & 0.1623 & 0.1389 & 0.5822 & 0.5254 & 0.8624 & 0.8405 & 0.4827 & 0.4485\\
\hline 
\end{tabular}
}
\end{table*}

Figure~\ref{fig:tsne2} visualizes the Task~1 feature distributions after sequential training through Tasks~1-4, providing a direct assessment of how well each method preserves previously learned representations during continual adaptation. FOCAL maintains a clearer and more structured separation between manipulated foreground and authentic background features than IFL and SAM+LoRA. Even after completing Task~4, the two feature clusters under FOCAL remain well separated, whereas SAM+LoRA exhibits substantial feature mixing and weaker class-discriminative structure. Quantitatively, FOCAL achieves the strongest final-stage separability on Task~1 among the three methods, with a larger inter-class feature distance than IFL and SAM+LoRA. These results suggest that FOCAL better preserves the discriminative structure learned from earlier tasks, which is crucial for retaining forgery localization capability throughout continual adaptation.


\subsection{Ablation Experiments}
\subsubsection{Effectiveness on Different Backbones}
To examine the flexibility of the proposed method for continual IFL, we further integrate FOCAL with alternative SAM backbones of different model capacities, including SAM-Tiny, SAM-Large, and the default SAM-Base. We report the evaluation results across different backbones under Protocol 1 in Table~\ref{tab:abl_backbone}. It can be observed that FOCAL consistently improves the overall performance across all three backbone variants and different stages of continual learning, demonstrating the adaptability of the proposed method. In particular, the performance gains become more evident as the learning sequence progresses, where the vanilla backbones exhibit substantial degradation on previously learned forgery domains after adapting to new tasks. In contrast, incorporating FOCAL generally maintains stronger performance on these previous domains while effectively adapting to newly introduced forgery data. This observation suggests that FOCAL provides a favorable balance between knowledge retention and continual adaptation. Overall, these results demonstrate that FOCAL can be seamlessly integrated with different SAM variants in a plug-and-play manner, highlighting its flexibility and generalizability for continual IFL.

\begin{table*}[t]
\centering
\caption{Impacts of the LoRA rank.}
\label{tab:abl_lora_rank}
\resizebox{\textwidth}{!}{
\begin{tabular}{c c cc cc cc cc cc}
\hline
\multirow{2}{*}{Setting} & \multirow{2}{*}{Task} 
& \multicolumn{2}{c}{Classic} 
& \multicolumn{2}{c}{DEFACTO} 
& \multicolumn{2}{c}{FantasticReality} 
& \multicolumn{2}{c}{TampCOCO} 
& \multicolumn{2}{c}{Average} \\
& 
& F1 & IoU 
& F1 & IoU 
& F1 & IoU 
& F1 & IoU 
& F1 & IoU \\
\hline					

\multirow{4}{*}{\begin{tabular}{c}rank = 4 \end{tabular}}
& T1 & 0.6399 & 0.5868  & -      & -      & -      & -      & -      & -      & 0.6399 & 0.5868 \\
& T2 & 0.3762 & 0.3256 & 0.3109 & 0.2704 & -      & -      & -      & -      & 0.3436 & 0.2980 \\
& T3 & 0.3417 & 0.2919 & 0.1608 & 0.1307 & 0.7051 & 0.6245 & -      & -      & 0.4025 & 0.3491 \\
& T4 & 0.3469 & 0.3053 & 0.1604 & 0.1343 & 0.6656 & 0.5950 & 0.8378 & 0.8099 & 0.5027 & 0.4611\\
\hline 

\multirow{4}{*}{\begin{tabular}{c}
rank = 8 \\
\end{tabular}}
& T1 & 0.6368 & 0.5871 & -      & -      & -      & -      & -      & -      & {0.6368} & {0.5871} \\
& T2 & 0.4286 & 0.3729 & 0.2921 & 0.2553 & -      & -      & -      & -      & {0.3604} & {0.3141}  \\
& T3 & 0.3578 & 0.3065 & 0.1700 & 0.1388 & 0.7075 & 0.6282 & -      & -      & {0.4117} & {0.3579} \\
& T4 & 0.3787 & 0.3319 & 0.1675 & 0.1417 & 0.6960 & 0.6264 & 0.8130 & 0.7834 & {0.5138} & {0.4708} \\
\hline

\multirow{4}{*}{\begin{tabular}{c}rank = 16\end{tabular}}
& T1 & 0.6785 & 0.6245  & -      & -      & -      & -      & -      & -      & \underline{0.6785} & 0.6245 \\
& T2 & 0.4158 & 0.3683 & 0.3392 & 0.2972 & -      & -      & -      & -      & 0.3775 & 0.3327 \\
& T3 & 0.3419 & 0.2965 & 0.1921 & 0.1562 & 0.6621 & 0.5819 & -      & -      & 0.3987 & 0.3448 \\
& T4 & 0.4038 & 0.3616 & 0.1724 & 0.1473 & 0.7252 & 0.6621 & 0.8696 & 0.8442 & 0.5427 & 0.5038\\
\hline 

\multirow{4}{*}{\begin{tabular}{c} rank = 32 \end{tabular}}
& T1 & 0.6801 & 0.6310  & -      & -      & -      & -      & -      & -      & \textbf{0.6801} & \underline{0.6310} \\
& T2 & 0.4813 & 0.4269 & 0.4086 & 0.3583 & -      & -      & -      & -      & \textbf{0.4450} & \textbf{0.3926} \\
& T3 & 0.3884 & 0.3399 & 0.2203 & 0.1811 & 0.6869 & 0.6057 & -      & -      & \underline{0.4319} & \underline{0.3756} \\
& T4 & 0.4386 & 0.3923 & 0.1970 & 0.1694 & 0.7761 & 0.7165 & 0.8771 & 0.8512 & \textbf{0.5722} & \textbf{0.5324}\\
\hline 

\multirow{4}{*}{\begin{tabular}{c} rank = 64 \end{tabular}}
& T1 & 0.6763 & 0.6331  & -      & -      & -      & -      & -      & -      & 0.6763 & \textbf{0.6331} \\
& T2 & 0.4721 & 0.4203 & 0.4023 & 0.3547 & -      & -      & -      & -      & \underline{0.4372} & \underline{0.3875} \\
& T3 & 0.4319 & 0.3802 & 0.2551 & 0.2082 & 0.7268 & 0.6477 & -      & -      & \textbf{0.4713} & \textbf{0.4120} \\
& T4 & 0.4192 & 0.3777 & 0.1552 & 0.1331 & 0.7346 & 0.6770 & 0.8906 & 0.8688 & \underline{0.5499} & \underline{0.5141}\\
\hline 
\end{tabular}
}
\end{table*}
\subsubsection{Effectiveness of the Designed Components}
FOCAL consists of the Forensic-Aware Representation Adaptation (FARA) and Fisher-weighted LoRA Gradient (FLAG) surgery modules, where FARA aims to adaptively extract robust forensic features, while FLAG is designed to preserve previously learned knowledge. We present the ablation results under Protocol 1 in Table~\ref{tab:module}. It can be observed that ablating either module leads to consistent performance degradation, demonstrating the effectiveness and complementarity of the two components. In particular, the removal of FARA results in an obvious overall performance drop, highlighting its importance in learning robust forensic representations and adapting them to evolving forgery distributions. Meanwhile, removing FLAG mainly degrades the performance on previously learned domains while maintaining competitive performance on the newly introduced domain. This observation suggests that FLAG effectively mitigates catastrophic forgetting by protecting knowledge important to previous tasks without sacrificing the plasticity required to learn emerging forgery patterns. Overall, these results demonstrate that FARA and FLAG jointly provide a favorable balance between learning robust forensic representations and preserving previously acquired knowledge, enabling FOCAL to effectively adapt to continually evolving forgery domains.

\subsubsection{Effectiveness of the Forensic Experts}
The ablation results in Table~\ref{tab:abl_moe} validate the effectiveness of the proposed Spatial Mixture-of-Forensic-Experts (SMoFE) module in continual image forgery localization. The full FOCAL model consistently achieves the best overall performance across the continual learning sequence, especially at the final stage T4, where it outperforms all expert-removal variants in terms of average F1 and IoU. This demonstrates that SMoFE effectively mitigates performance degradation during continual adaptation by dynamically integrating complementary forensic cues. Removing any expert weakens the model’s overall ability to mine robust forensic traces, with the removal of Bayer and SRM experts causing particularly large drops. These results indicate that SMoFE can adaptively capture diverse low-level forensic evidence, including demosaicing artifacts, noise residuals, boundaries, and compression traces, thereby enhancing both forensic feature mining and continual learning performance.


\subsubsection{Ablation Study on LoRA Rank}

Table~\ref{tab:abl_lora_rank} reports the effect of the LoRA rank on continual image forgery localization. Small ranks, such as 4 and 8, provide limited adaptation capacity and yield weaker performance when learning newly introduced forgery domains. Increasing the rank generally improves continual adaptation, with rank 32 achieving the best overall performance at the final stage T4. This indicates that rank 32 provides sufficient capacity to learn new manipulation patterns while retaining previously acquired knowledge. Although rank 64 remains competitive, its lower final performance suggests that further increasing the adaptation capacity provides limited benefits, introduces redundant parameters, and may increase the risk of overfitting. We therefore adopt rank 32 as the default setting in all experiments.


    



\section{Conclusion}
To the best of our knowledge, we present the first continual learning framework for Image Forgery Localization (IFL), aiming to enable IFL models to adapt to sequentially arriving forgery data while preserving previously acquired forensic knowledge. We first establish a comprehensive benchmark with two realistic data evolution protocols, namely, cross-dataset and cross-content protocols, to systematically evaluate the continual adaptation capabilities of existing IFL and continual learning methods. Experimental results reveal that existing methods suffer from severe performance degradation when adapting to emerging forgery domains, particularly in the presence of substantial domain gaps.

To address these challenges, we propose FOrensic-aware Continual Adaptation with LoRA surgery (FOCAL), which comprises a forensic-aware representation adaptation module and a Fisher-weighted LoRA gradient surgery strategy. 
Specifically, the Spatial Mixture-of-Forensic-Experts (SMoFE) mechanism adaptively integrates complementary forensic cues from multiple trace experts, enabling the model to capture robust manipulation evidence across diverse and previously unseen domains. Forensic Evidence-Guided Dense Prompting (FEGDP) bridges the forensic-to-prompt representation gap by converting low-level artifacts into structured localization evidence for SAM.
Meanwhile, Fisher-weighted LoRA Gradient (FLAG) surgery selectively modulates the gradients of LoRA parameters according to their importance to previously learned tasks, thereby mitigating catastrophic forgetting while retaining sufficient plasticity to learn new forgery patterns. 
Extensive quantitative and visualization experiments under both protocols demonstrate that FOCAL consistently achieves superior performance in both pixel-level forgery localization and image-level forgery detection. Despite considering two practical data evolution protocols, the proposed benchmark does not fully capture the complexity of real-world continual learning scenarios. Future work could therefore incorporate more complex learning protocols and more diverse datasets to further advance continual learning for real-world multimedia forensics. We hope that this work provides a strong benchmark and baseline for future research in this important direction.

\section*{Declarations}
\textbf{Data Availability Statements.} 
We conducted experiments using several publicly accessible image forgery localization datasets: 
CASIAv2 \citep{Dong2013}, IFC \citep{IFC}, WildWeb \citep{zampoglou2015detecting}, IMD2020 \citep{novozamsky2020imd2020}, DEFACTO \citep{mahfoudi2019defacto}, FantasticReality \citep{kniaz2019point}, TampCOCO \citep{kwon2022learning}, RSIID \citep{cardenuto2022benchmarking}, 
Biofors \citep{sabir2021biofors}, 
SciSp \citep{lin2024exposing},  
STFD \citep{yu2023learning}, and  
DocTamper \citep{qu2023towards}. These datasets are critical to our analysis and are readily available online. Specifically, 
CASIAv2 is accessible at \url{https://github.com/namtpham/casia2groundtruth};
IFC is accessible at \url{https://signalprocessingsociety.org/newsletter/2013/06/ifs-tc-image-forensics-challenge};
WildWeb is accessible at \url{https://mklab.iti.gr/results/the-wild-web-tampered-image-dataset/};
IMD2020 is accessible at \url{https://staff.utia.cas.cz/novozada/db/};
DEFACTO is accessible at \url{https://defactodataset.github.io/};
FantasticReality is accessible at \url{https://github.com/mjkwon2021/CAT-Net/issues/51#issuecomment-2537517937};
TampCOCO is accessible at \url{https://www.kaggle.com/datasets/qsii24/tampcoco};
RSIID is accessible at \url{https://zenodo.org/records/15095089};
Biofors is accessible at \url{https://drive.google.com/file/d/1UVSJ6h7r8pmOWYZkqWeAZ_YvwbFr1wV3/view};
SciSp is accessible at \url{https://zenodo.org/records/10989921};
STFD is accessible at \url{https://drive.google.com/file/d/12MvM1R4R2iE-KGZmB1BSAXqR1Wo1kPIj/view};
DocTamper is accessible at \url{https://github.com/qcf-568/DocTamper}.

\bibliography{main}%
\end{document}